\documentclass[10pt]{article}

\usepackage{amsmath}

\usepackage{array}

\usepackage{appendix}

\usepackage{tocloft}                   

\usepackage{graphicx}

\usepackage{amsfonts}

\usepackage{amssymb}

\usepackage{undertilde}

\usepackage{mathrsfs}

\usepackage{yfonts}

\usepackage{yhmath}                  

\usepackage{euscript}

\usepackage{centernot}                 

\usepackage{ifsym}                     

\usepackage{lmodern}                   

\usepackage{upgreek}

\usepackage{mathtools}

\usepackage{mathabx}

\usepackage{xcolor}

\usepackage{slantsc}
\usepackage{calligra}

\usepackage{bbold}          

\usepackage[T1]{fontenc}

\usepackage{epsf}

\usepackage{latexsym}

\usepackage{tipa}          

\usepackage{makeidx}

\makeindex

\def\b{\begin{equation}}

\def\e{\begin{equation}}

\def\be{\begin{equation}}              

\def\ee{\end{equation}}

\def\beq{\begin{equation}}

\def\eeq{\end{equation}}

\def\bea{\begin{eqnarray}}

\def\eea{\end{eqnarray}}

\def\m{\mbox{ }}

\def\mma {\m , \m \m }

\def\!{\hspace{-1.6667em}}

\def\n{\noindent}

\def\u{\underline}       

\def\uc{\underbracket}   
\def\uo{\utilde}         

\def\w{\widetilde}

\def\slLambda{\mathit{\Lambda}}                   

\def\bip{\mbox{\boldmath$p$}}

\def\biq{\mbox{\boldmath$q$}}

\def\biu{\mbox{\boldmath$u$}}

\def\bea{\mbox{\boldmath\ttfamily{B}}}

\def\biC{\mbox{\boldmath$C$}}              

\def\biD{\mbox{\boldmath$D$}}

\def\sbiG{\mbox{\boldmath \scriptsize $G$}} 

\def\biL{\mbox{\boldmath$L$}}

\def\biM{\mbox{\boldmath$M$}}

\def\biN{\mbox{\boldmath$N$}}

\def\biO{\mbox{\boldmath$O$}}

\def\biP{\mbox{\boldmath$P$}}

\def\biQ{\mbox{\boldmath$Q$}}

\def\biW{\mbox{\boldmath$W$}}

\def\biX{\mbox{\boldmath$X$}}

\def\sbiC{\mbox{\ttfamily\fontseries{b}\selectfont C}} 

\def\sbiG{\mbox{\ttfamily\fontseries{b}\selectfont G}}

\def\sbiD{\mbox{\ttfamily\fontseries{b}\selectfont D}} 

\def\sbiQ{\mbox{\scriptsize\boldmath$Q$}}

\def\birho{\mbox{\boldmath$\rho$}}

\def\brho{\birho}                                   

\def\bxi{\mbox{\boldmath$\xi$}} 

\def\bchi{\mbox{\boldmath$\chi$}} 

\def\bzeta{\mbox{\boldmath$\zeta$}} 

\def\mA{\mbox{A}}  

\def\mB{\mbox{B}}  

\def\mE{\mbox{E}}                        

\def\mG{\mbox{G}}

\def\mJ{\mbox{J}}

\def\mL{\mbox{L}}

\def\mM{\mbox{M}}                        

\def\mN{\mbox{N}}

\def\mP{\mbox{P}}

\def\mQ{\mbox{Q}}

\def\mS{\mbox{S}}                        

\def\ma{\mbox{a}}

\def\mb{\mbox{b}}

\def\mf{\mbox{f}}

\def\mg{\mbox{g}}

\def\mh{\mbox{h}}

\def\mp{\mbox{p}}

\def\bP{\mbox{\bf P}}

\def\bQ{\mbox{\bf Q}}

\def\bdelta{\mbox{\boldmath$\delta$}}

\def\bsigma{\mbox{\boldmath$\sigma$}}                   %

\def\bupSigma{\mbox{\boldmath$\Sigma$}}                 

\def\scM{\mbox{\scriptsize ${\cal M}$}}

\def\fD{\mbox{\sffamily D}}

\def\fN{\mbox{\sffamily N}}

\def\fO{\mbox{\sffamily O}}

\def\fP{\mbox{\sffamily P}}

\def\fQ{\mbox{\sffamily Q}}

\def\sh{\mbox{\scriptsize h}}

\def\sp{\mbox{\scriptsize p}}

\def\sw{\mbox{\scriptsize w}}

\def\sA{\mbox{\scriptsize A}}

\def\sE{\mbox{\scriptsize E}}

\def\sL{\mbox{\scriptsize L}} 

\def\sM{\mbox{\scriptsize M}}

\def\sS{\mbox{\scriptsize S}}

\def\sW{\mbox{\scriptsize W}}

\def\sY{\mbox{\scriptsize Y}}

\def\sfx{\mbox{\sffamily{\scriptsize x}}}     

\def\sfA{\mbox{\sffamily{\scriptsize A}}}     

\def\sfB{\mbox{\sffamily{\scriptsize B}}}     

\def\sfC{\mbox{\sffamily{\scriptsize C}}}     

\def\sfD{\mbox{\sffamily{\scriptsize D}}}      

\def\sfE{\mbox{\sffamily{\scriptsize E}}}      

\def\sfN{\mbox{\sffamily{\scriptsize N}}}      

\def\sfO{\mbox{\sffamily{\scriptsize O}}}      

\def\sfP{\mbox{\sffamily{\scriptsize P}}}      

\def\sfQ{\mbox{\sffamily{\scriptsize Q}}}      

\def\sfR{\mbox{\sffamily{\scriptsize R}}}      

\def\sfV{\mbox{\sffamily{\scriptsize V}}}      

\def\sfW{\mbox{\sffamily{\scriptsize W}}}      

\def\sttV{\mbox{\scriptsize ${\ttV}$}}

\def\sttW{\mbox{\scriptsize ${\ttW}$}}

\def\sbw{\mbox{{\bf \scriptsize w}}}

\def\sbS{\mbox{{\bf \scriptsize S}}}

\def\sbcC{\mbox{\boldmath \scriptsize ${\cal C}$}}

\def\sbcF{\mbox{\boldmath \scriptsize ${\cal F}$}}

\def\sbcG{\mbox{\boldmath \scriptsize ${\cal G}$}}

\def\sbcS{\mbox{\boldmath \scriptsize ${\cal S}$}}

\def\tiV{\mbox{\tiny $V$}}

\def\tiW{\mbox{\tiny $W$}}

\def\bfQ{\mbox{{\bf \sffamily Q}}}                                    

\def\bfP{\mbox{{\bf \sffamily P}}}                                    

\def\sbcC{\mbox{{\boldmath \scriptsize${\cal C}$}}}                               

\def\ttV{\mbox{\tt V}}

\def\ttW{\mbox{\tt W}}

\def\bttu{\mbox{\boldmath {\tt U}}}

\def\bttU{\mbox{\boldmath {\tt U}}}

\def\siQ{\mbox{\scriptsize $Q$}}

\def\cr{\mbox{\scriptsize{\bf $\m  \times \m $}}}

\def\sumi2{\sum\mbox{}_{\mbox{}_{\mbox{\scriptsize $i$=1}}}^2}

\def\sumi3{\sum\mbox{}_{\mbox{}_{\mbox{\scriptsize $i$=1}}}^3}

\def\sumA{\sum\mbox{}_{\mbox{}_{\mbox{\sfA}}}}

\def\sumABcycles3{\sum\mbox{}_{\mbox{}_{\mbox{\scriptsize cycles $A,B$=1}}}^{3}}

\def\sumCDcycles3{\sum\mbox{}_{\mbox{}_{\mbox{\scriptsize cycles $C,D$=1}}}^{3}}

\def\sumIN{\sum\mbox{}_{\mbox{}_{\mbox{\scriptsize $I$=1}}}^{N}}

\def\sumj3{\sum\mbox{}_{\mbox{}_{\mbox{\scriptsize $j$=1}}}^3}

\def\sumk3{\sum\mbox{}_{\mbox{}_{\mbox{\scriptsize $k$=1}}}^3}

\def\prodiA1{\prod\mbox{}_{\mbox{}_{\mbox{\scriptsize $i$=1}}}^{A - 1}}

\def\d{\textrm{d}}                                                  

\def\pa{\partial}                                                   

\def\partional{\bdelta\hspace{-0.08in}\pa \, }                          

\def\es{\m = \m}

\def\:={\m := \m}

\def\=:{\m =: \m}

\def\Mom{\mbox{{\boldmath$\mathfrak{P}$}}}                     

\def\FrT{\mathfrak{T}}                                         

\def\FrC{\mbox{$\mathfrak{C}$}}                                
                                                       
\def\FrU{\mbox{$\mathfrak{U}$}}                                
\def\FrD{\mbox{$\mathfrak{D}$}}	                               
 
\def\FrM{\mbox{$\mathfrak{M}$}}                                
\def\lFrg{\mbox{\Large$\mathfrak{g}$}}                         
\def\FrH{\mbox{$\mathfrak{H}$}}                                
\def\FrF{\mbox{\boldmath$\mathfrak{F}$}}                       

\def\FrT{\mbox{\boldmath$\mathfrak{T}$}}                       
\def\FrG{\mathfrak{G}}                                         
\def\sFrG{\mbox{\boldmath\scriptsize$\mathfrak{G}$}}           

\def\TruePhase{\mbox{$\FrT$rue-$\Phase$}}   

\def\Hilb{\mbox{{\boldmath$\mathfrak{H}$}ilb}}                 
\def\bFrW{\mbox{\boldmath$\mathfrak{W}$}} 					   

\def\CanObs{\FrC\mbox{an-}\FrO\mbox{bs}}                                

\def\CanUnresObs{\FrC\mbox{an-}\FrU\mbox{nres-}\FrO\mbox{bs}}                                

\def\GeomUnresObs{\FrG\mbox{eom-}\FrU\mbox{nres-}\FrO\mbox{bs}}                                

\def\MomUnresObs{\FrM\mbox{om-}\FrU\mbox{nres-}\FrO\mbox{bs}}

\def\DiracObs{\FrD\mbox{irac-}\FrO\mbox{bs}}                                

\def\GeomGaugeObs{\FrG\mbox{eom-}\FrG\mbox{auge-}\FrO\mbox{bs}}

\def\MomGaugeObs{\FrM\mbox{om-}\FrG\mbox{auge-}\mbox{bs}}

\def\CanGaugeObs{\FrC\mbox{an-}\FrG\mbox{auge-}\FrO\mbox{bs}}

\def\ChronosObs{\FrC\mbox{hronos-}\FrO\mbox{bs}}                             %

\def\scC{\mbox{\scriptsize ${\cal C}$}}                    
\def\scG{\mbox{\scriptsize ${\cal G}$}}                    

\def\scH{\mbox{\scriptsize ${\cal H}$}}                    

\def\scQ{\mbox{\scriptsize ${\cal Q}$}}                    

\def\scS{\mbox{\scriptsize ${\cal S}$}}                    

\def\bFlin{\sbcF\mbox{\bf lin}} 

\def\Quad{\scQ\mbox{uad}}                                  

\def\Chronos{\scC\mbox{hronos}}                            

\def\Gauge{\sbcG\mbox{auge}}

\def\bGauge{\sbcG\mbox{\bf auge}} 

\def\bShuffle{\sbcS\mbox{\bf huffle}} 

\def\observables{\mbox{\ttfamily\fontseries{b}\selectfont B}} 

\def\chronos{\mbox{\ttfamily\fontseries{b}\selectfont C}} 
					   
\def\bttD{\mbox{\ttfamily\fontseries{b}\selectfont D}}     %

\def\bttA{\mbox{\ttfamily\fontseries{b}\selectfont A}}     %

\def\bttO{\mbox{\ttfamily\fontseries{b}\selectfont O}}     %
														
\def\bttV{\mbox{\ttfamily\fontseries{b}\selectfont V}} 
											
\def\bttW{\mbox{\ttfamily\fontseries{b}\selectfont W}} 
								
\def\bttC{\mbox{\ttfamily\fontseries{b}\selectfont C}}

\def\bttG{\mbox{\ttfamily\fontseries{b}\selectfont G}}

\def\Dirac{\mbox{\ttfamily\fontseries{b}\selectfont D}}    
														   
\def\gauge{\mbox{\ttfamily\fontseries{b}\selectfont G}}                  

\def\Kuchar{\mbox{\ttfamily\fontseries{b}\selectfont K}}                  

\def\Gaugeo{\mbox{\ttfamily\fontseries{b}\selectfont G}}                  
                                      
\def\unres{\mbox{\ttfamily\fontseries{b}\selectfont U}}                   %

\def\unres{\mbox{\ttfamily\fontseries{b}\selectfont U}}
                                 
\def\FrQ{\mbox{\Large $\mathfrak{q}$}}                               
\def\Phase{\mbox{{\boldmath$\mathfrak{P}$}hase}}                     

\def\bFrR{\mbox{\boldmath$\mathfrak{R}$}}                            
\def\Rig-Phase{\bFrR\mbox{ig-}\Phase}                                
                                                                													   
\def\FrP{\mbox{\Large $\mathfrak{p}$}}                                 
\def\bFrM{\mbox{\boldmath${\mathfrak{M}}$}}

\def\bFrR{\mbox{\boldmath$\mathfrak{R}$}}                            

\def\bFrR{\mbox{\boldmath$\mathfrak{R}$}}                            

\def\1mat{\u{\u{1}}}                                                 

\def\lFrs{\mathfrak{S}}                                              

\def\Positive-Modespace{\mbox{{\boldmath$\mathfrak{M}$}odespace$^+$}}

\def\POSITIVE-MODESPACE{\mbox{{\boldmath$\mathfrak{M}$}ODESPACE$^+$}}

\def\FrO{\mbox{$\mathfrak{O}$}}                                      

\def\Kin-Hilb{\mbox{{\boldmath$\mathfrak{K}$}in-\Hilb}}                     

\def\Mid-Hilb{\mbox{{\boldmath$\mathfrak{M}$}id-\Hilb}}                     

\def\Dyn-Hilb{\mbox{{\boldmath$\mathfrak{D}$}yn-\Hilb}}                     

\def\5Star{\mbox{\Large$\star$}}              

\def\K{Kucha\v{r} }

\def\peq{\m \mbox{`='} \m}

\begin{document}

\begin{center}

\large{\bf A LOCAL RESOLUTION OF THE PROBLEM OF TIME}

\vspace{.1in}

\large{\bf VIII. Assignment of Observables}

\vspace{.1in}

{\bf Edward Anderson}$^1$ 

\vspace{.1in}

{\bf \it based on calculations done at Peterhouse, Cambridge} 

\end{center}

\begin{abstract}

Given a state space, Assignment of Observables involves Taking Function spaces Thereover. 
At the classical level, the state space in question is phase space or configuration space.
This assignment picks up nontrivialities when whichever combination of constraints and the quantum apply.

For Finite Theories, weak observables equations are inhomogeneous-linear first-order PDE systems. 
Their general solution thus splits into complementary function plus particular integral: strong and nontrivially-weak observables respectively. 
We provide a PDE analysis for each of these. 
In the case of single observables equations -- corresponding to single constraints -- the Flow Method readily applies.
Finding all the observables requires free characteristic problems.
For systems, this method can be applied sequentially, due to integrability conferred by Frobenius' Theorem. 
In each case, the first part of this approach is Lie's Integral Theory of Geometrical Invariants, or the physical counterpart thereof. 
The second part finds the function space thereover, giving the entire space of (local) observables.

We also outline the Field Theory counterpart. 
Here one has functional differential equations. 
Banach (or tame Fr\'{e}chet) Calculus is however sufficiently standard for the Flow Method and free characteristic problems to still apply. 
These calculi support the Lie-theoretic combination of machinery that our Local Resolution of the Problem of Time requires, 
by which Field Theory and GR are included.   

\end{abstract}

$^1$ dr.e.anderson.maths.physics *at* protonmail.com   

\section{Introduction}\label{Intro}

This is the eighth Article \cite{I, II, III, IV, V, VI, VII} on A Local Resolution of the Problem of Time 
\cite{Battelle, DeWitt67, Dirac, K81, K91, K92, I93, K99, APoT, FileR, APoT2, AObs, APoT3, ALett, ABook, A-CBI} 
and its underlying Local Theory of Background Independence. 
We here expand on Sec 3 of Article III's opening account of Assignment of Observables \cite{DiracObs, K92, I93, K93}, now atop \cite{AObs, ABook} 
the triple unification of                                   Constraint Closure        \cite{Dirac51, Dirac58, Dirac, SBook, HTBook, ABook, III, VII} 
with                                                        Temporal                  \cite{BSW, BB82, FileR, TRiPoD, ABook, I, V, VI}             
and                                                         Configurational           \cite{ADM, BB82, FileR, ABook, II, V, VI}                    Relationalism.  
That we can treat this after completing the triple, 
and separately from Article IX's Constructability extension, 
is one of the great decouplings of Problem of Time Facets \cite{ABook, I}. 
Resolving the triple of facets gives a consistent phase space, $\Phase$. 
Taking Function Spaces {\sl Thereover} -- the essence of Assigning Observables -- entails addressing a {\sl subsequent} mathematical problem {\sl on} $\Phase$,  
rather than imposing some further conditions on whether that $\Phase$ is adequate. 

\m 

\n We pose concrete mathematical problems for Finite Theory's observables at the level of brackets algebras in Sec  \ref{Bra} 
                                                                                   and of PDEs              in Secs \ref{S-PDE} and \ref{W-PDE}. 
Weak observables equations are inhomogeneous-linear first-order PDE systems. 
Their general solution thus splits into complementary function plus particular integral: strong and nontrivially-weak observables respectively. 
In the case of single observables equations -- corresponding to single constraints -- the {\bf Flow Method} readily applies.
Finding all the observables requires {\bf free characteristic problems}.
For {\bf systems} -- corresponding to multiple constraints -- this approach can be applied sequentially, 
due to {\bf integrability} as conferred by {\bf Frobenius' Theorem} \cite{AMP, Lee2}.   
In each case, the first part of our approach is {\bf Lie's Integral Theory of Geometrical Invariants} (or some physical counterpart thereof). 
The second part is Finding the Function space Thereover, giving the entire space of (local) observables. 
The more practical matter of Expression in Terms of Observables -- requiring only enough observables to span phase (or configuration) space -- 
is covered in Sec \ref{EiToO}.
Strategies for addressing the Problem of Observables are in Sec \ref{Strat}. 
We also give explicit examples of quite a number of distinct notions of observables \cite{DiracObs, HTBook, K92, I93, K93, AObs, ABook, PE-1, DO-1} 
                                                                      in Sec \ref{Solns}. 
Purely geometrical such moreover play a further role \cite{PE-1, PE-2-3} in the Foundations of Geometry \cite{Hilb-Ax, HC32, Coxeter, S04, Stillwell, PE-1}, 
as well as coinciding with the configuration space $\FrQ$ restriction of $\Phase$.  

\m 

\n The Field Theory and GR counterpart is in Sec \ref{F-Obs}. 
This now gives FDEs -- functional differential equations -- as its observables equations. 
Similar Flow Method and sequential use of free characteristic problems carry over, however,   
thanks to the underlying benevolence of {\bf Banach Calculus} \cite{AMP} (for now, or tame Fr\'{e}chet Calculus \cite{Hamilton82} more generally).
These calculi support the Lie-theoretic combination of machinery required by our Local Resolution of the Problem of Time,   
by which Field Theory and GR are included.

\m 

\n Appendix \ref{TRi-Obs} keeps track of TRi (Temporal Relationalism implementing) modifications 
                                   and other Problem of Time facet interferences in part involving Assignment of Observables.  

%
%
%
%
%
%
%
%
%
%
%
%
%
%
%
%
%
%
%

\section{Brackets-level considerations}\label{Bra}

\subsection{Strong observables--constraints system} 

{\bf Structure 1} Strong observables $\bttO$ extend the constraint algebraic structure $\FrC(\lFrs)$ as follows. 
\be 
\mbox{\bf \{} \uc{\sbcC} \mbox{\bf ,} \, \uc{\sbcC} \mbox{\bf \}}  \es  \uc{\uc{\uc{\biC}}} \, \uc{\sbcC}  \m ,
\ee 
\be 
\mbox{\bf \{} \uc{\sbcC} \mbox{\bf ,} \, \uo{ \bttO} \mbox{\bf \}}  \es            0                       \m , 
\label{S-Obs-Def}
\ee 
\be 
\mbox{\bf \{} \uo{\bttO} \mbox{\bf ,} \, \uo{\bttO} \mbox{\bf \}} \es \uo{\uo{\uo{\biO}}} \, \uo{\bttO}    \m . 
\ee 
{\bf Remark 1} This is overall \cite{ABook} of direct product form,   
\be 
\FrC(\lFrs)  \m \times \m  \CanObs(\lFrs)                                                                             \m .  
\ee

\subsection{Weak observables--constraints system}

{\bf Structure 2} Weak observables $\bttO^{\sw}$ extend the constraint algebraic structure as follows. 
\be 
\mbox{\bf \{} \uc{\sbcC}       \mbox{\bf ,} \, \uc{\sbcC}       \mbox{\bf \}}  \es  \uc{\uc{\uc{\biC}}} \, \uc{\sbcC}        \m ,
\label{C-C}
\ee 
\be 
\mbox{\bf \{} \uc{\sbcC}       \mbox{\bf ,} \, \uo{\bttO}^{\sw} \mbox{\bf \}}  \es  \uc{\uo{\uc{\biW}}} \, \uc{\sbcC}        \m , 
\label{W-Obs-Def}
\ee 
\be 
\mbox{\bf \{} \uo{\bttO}^{\sw} \mbox{\bf ,} \, \uo{\bttO}^{\sw} \mbox{\bf \}}  \es  \uo{\uo{\uo{\biO}}} \, \uo{\bttO}^{\sw}  \m .  
\ee 
\n{\bf Remark 1} The second equation signifies that $\sbcC$ is a good $\CanObs(\lFrs)$-object. 
This is a consequence of constraints closing weakly, via the Jacobi identity.  

\m 

\n{\bf Remark 2} This is now overall \cite{ABook} of semidirect product form,   
\be 
\FrC(\lFrs) \m \rtimes \m \CanObs(\lFrs)   \m .  
\ee 
\n{\bf Remark 3} Comparing (\ref{C-C}) and (\ref{S-Obs-Def}) or (\ref{W-Obs-Def}) implies that the $\sbcC$ are themselves in some sense observables.
However, since we already knew that $\sbcC \approx 0$, 
in studying observables we are really looking for further quantities outside of this trivial case.  
Let us call these other quantities {\it proper observables}; the rest of the Series will always take `observables' to mean this.  
 
\m 

\n{\bf Definition 1} We term the general solution of the weak observables equation $\bttO^{\sW}$ {\it weak observables}. 

\m 

\n{\bf Remark 4} The weak observables equation is moreover an inhomogeneous counterpart of the homogeneous linear strong observables equation. 
So 
\be 
\bttO^{\sW} = \bttO + \bttO^{\sw} 
\ee 
in the manner of a complementary function plus particular integral split of the general solution of an inhomogeneous linear equation, 
\be 
\mbox{GS} = \mbox{CF} + \mbox{PI}                                                                                          \m .  
\ee 
{\bf Definition 2} We term particular-integral solutions of the weak equation $\bttO^{\sw}$ 
-- weak observables which are explicitly independent of any strong observables -- {\it nontrivially weak observables}.  

\m 

\n{\bf Remark 5} The pure configurational geometry case cannot support any nontrivially-weak observables. 
This is because we are in a context in which constraints have to depend on momenta
\be 
\sbcC = \sbcC(\biQ, \, \biP) \m \mbox{ \sl with specific $\biP$ dependence }                                               \m .    
\ee
A fortiori, first-class linear constraints refer specifically to being linear in their momenta $\biP$, 
By this, the weak configurational observables equation would have a momentum-dependent inhomogeneous term right-hand side. 
But this is inconsistent with admitting a solution with purely $\biQ$-dependent right-hand-side.

\subsection{Why not a more general $\sbcC$, $\biO^{\sbw}$ system?}

\n Na\"{\i}ve algebraic generality suggests the more general form  
\be 
\mbox{\bf \{} \uc{\sbcC}       \mbox{\bf ,} \, \uc{\sbcC}       \mbox{\bf \}}   \es  \uc{\uc{\uc{\biC}}}              \, \uc{\sbcC}        \m + \m  
                                                                                     \uc{\uc{\uo{\biD}}}              \, \uo{\bttO}^{\sw}  \m ,
\ee 
\be 
\mbox{\bf \{} \uc{\sbcC}       \mbox{\bf ,} \, \uo{\bttO}^{\sw} \mbox{\bf \}}   \es  \uc{\uc{\uc{\biW}}}              \, \uc{\sbcC}        \m + \m  
                                                                                     \uc{\uo{\uo{\biX}}}              \, \uo{\bttO}^{\sw}  \m , 
\ee 
\be 
\mbox{\bf \{} \uo{\bttO}^{\sw} \mbox{\bf ,} \, \uo{\bttO}^{\sw} \mbox{\bf \}}   \es  \uo{\uo{\uc{\biL}}}              \, \uc{\sbcC}        \m + \m  
                                                                                     \uo{\uo{\uo{\biO}}}\mbox{}^{\sw} \, \uo{\bttO}^{\sw}  \m . 
\ee
\n However, $\biD = 0$ since physical systems provide the constraints $\sbcC$ without reference to observables $\bttO$.  

\m 

\n $\biX = 0$ preserves $\sbcC$'s status as a good $\CanObs(\lFrs)$-object.   

\m 

\n $\biL \neq 0$ gives a further sense of weak, though it prevents the observables from forming a subalgebra. 
Now starting with the observables would imply the constraints, but the constraints are already prescribed elsewise by the physics.

\subsection{Relation between $\biO^{\sw}$ and $\biW$ structure constants}

{\bf Lemma 1} 
\be 
\left(  2 \, {W^{\sfB}}_{\sfA[\sfO} \, {\delta_{\sfP]}}^{\sfQ} - 
             {\delta_{\sfA}}^{\sfB} {O^{\sw \, \sfQ}}_{\sfO\sfP} \right) {W^{\sfC}}_{\sfB\sfQ} \, \scC_{\sfC}  \es  0
\label{factors}
\ee
\n{\u{Proof}} 
\be 
0  \es  J(\sbcC, \, \bttO^{\sw}, \, \bttO^{\sw}) 
   \es  \left(  
                 {W^{\sfC}}_{\sfA\sfO} \, {\delta_{\sfC}}^{\sfD} {\delta_{\sfP}}^{\sfO} - 
				 {W^{\sfC}}_{\sfA\sfP} \, {\delta_{\sfC}}^{\sfD} {\delta_{\sfO}}^{\sfR} - {O^{\sw \, \sfQ}}_{\sfO\sfP} \, {\delta_{\sfA}}^{\sfD} {\delta_{\sfQ}}^{\sfR}  
		\right)  {W^{\sfE}}_{\sfD\sfR} \,  \scC_{\sfE}   
\ee
and factorize. $\Box$ 

\m 

\n{\bf Corollary 1} This can be achieved by 
\be 
\biW = 0  \m , 
\ee
returning the strong case,    or by 
\be 
{O^{\sw \, \sfQ}}_{\sfO\sfP}  \es  \frac{2}{c} \, {W^{\sfA}}_{\sfA[\sfO}{\delta_{\sfP]}}^{\sfQ}  
\ee
with $c := \mbox{dim}(\FrC)$, or by a zero double-trace `perpendicularity' condition.  

\m 

\n{\bf Remark 1} On the other hand, no capacity to influence $\biO^{\sw}$ can be found in 
\be 
J(\sbcC, \sbcC, \bttO^{\sw}) = 0 
\ee

\subsection{Strong and weak observables considered together}

\n{\bf Structure 3} By our CF + PI split, the strong and nontrivially-weak canonical observables algebra is consequently of the direct product form 
\be 
\CanObs^{\sW}(\lFrs)  \es  \CanObs^{\sw}(\lFrs) \times \CanObs(\lFrs)   \m .   
\ee
\n{\bf Structure 4} Including the constraints as well, we have the 3-block algebraic structure. 
\be 
\mbox{\bf \{} \uc{\sbcC} \mbox{\bf ,} \, \uc{\sbcC} \mbox{\bf \}}               \es  \uc{\uc{\uc{\biC}}} \, \uc{\sbcC}             \m ,
\ee 
\be 
\mbox{\bf \{} \uc{\sbcC} \mbox{\bf ,} \, \uo{\bttO} \mbox{\bf \}}               \es            0                                   \m , 
\ee 
\be 
\mbox{\bf \{} \uo{\bttO} \mbox{\bf ,} \, \uo{\bttO} \mbox{\bf \}}               \es \uo{\uo{\uo{\biO}}} \, \uo{\bttO}              \m , 
\ee 
\be 
\mbox{\bf \{} \uc{\sbcC} \mbox{\bf ,} \, \uo{\bttO}^{\sw} \mbox{\bf \}}         \es  \uc{\uo{\uc{\biW}}} \, \uc{\sbcC}             \m , 
\ee 
\be 
\mbox{\bf \{} \uo{\bttO^{\sw}} \mbox{\bf ,} \, \uo{\bttO}^{\sw} \mbox{\bf \}}  \es  \uo{\uo{\uo{\biO}}}^{\sw} \, \uo{\bttO}^{\sw}  \m ,  
\ee 
\be 
\mbox{\bf \{} \uo{\bttO} \mbox{\bf ,} \, \uo{\bttO}^{\sw} \mbox{\bf \}}        \es  0                                              \m .  
\ee
\n{\bf Remark 1} This includes each of $\sbcC$, $\bttO$ and $\bttO^{\sw}$ as subalgebras, by the first, third and fifth equations respectively. 

\m

\n{\bf Remark 2} The sixth equation's zero right-hand-side is part of the implementation of the CF to PI linear independence.  
This is partly enforced by 
\be 
J(\sbcC, \bttO, \bttO^{\sw}) = 0  \m ,
\ee 
which renders an initial $\bttO$ and $\bttO^{\sw}$ combination pure-$\bttO$.

\m 

\n{\bf Remark 3} The overall algebraic structure is of the form 
\be 
\left( \FrC(\lFrs) \rtimes \CanObs^{\sw}(\lFrs) \right) \times \CanObs(\lFrs)   \m .   
\ee

\section{Strong observables PDEs}\label{S-PDE}

One can obtain explicit PDEs by writing out what the Poisson brackets definition of constrained observables means \cite{AObs2}.
For strong observables, (\ref{S-Obs-Def}) gives  
\be 
0  \es  \mbox{\bf \{} \sbcC \mbox{\bf ,} \, \bttO \mbox{\bf \}}  
   \es  \frac{\pa \,  \sbcC}{\pa \,  \biQ}\frac{\pa \,  \bttO}{\pa \,  \biP} - \frac{\pa \,  \sbcC}{\pa \,  \biP} \frac{\pa \,  \bttO}{\pa \,  \biQ}         \m .
\ee
We can moreover take 
\be 
\frac{\pa \,  \sbcC}{\pa \,  \biQ} \m \mbox{ and } \m \frac{\pa \,  \sbcC}{\pa \,  \biP} \m \mbox{ to be knowns}  \m , 
\ee 
leaving us with a homogeneous-linear first-order PDE system. 
I.e.\ a homogeneous subcase 
\be 
\sum_{A} a^{A}(x^{B}, \phi) \pa_A \phi  \es  0
\ee 
of the claimed first-order linear form (III.87). 
First order their containing first-order partial derivatives $\pa_{\alpha}\phi$ and no higher.  
Linear refers to the unknown variables $\bttO$.  

\m 

\n{\bf Remark 1} As the strong case involves a homogeneous equation, 
\be
\bttO = \mbox{const}   \m ,
\ee
is always a solution.   
We refer to this as the {\it trivial solution}.
We call all other solutions of first-order homogeneous quasilinear PDEs {\it proper solutions}: a nontrivial kernel condition.

\subsection{Single strong observables equation}

The Flow Method \cite{John, Lee2} immediately applies here.

\m 

\n{\bf Structure 5} This gives a corresponding ODE system of form 
\be 
\dot{x}^A = a^A(x)                 \m , 
\label{x-dot}
\ee  
\be 
\dot{\bttO} = 0                    \m . 
\label{phi-dot}
\ee
Here, the dot denotes  
\be
\frac{\d}{\d \nu}  \m , 
\ee
for $\nu$ a fiducial variable to be eliminated, rather than carrying any temporal (or other geometrical or physical) significance. 

\m 

\n{\bf Remark 1} In the geometrical setting, the first equation here corresponds to Lie's Integral Approach to Geometrical Invariants.  

\m 

\n{\bf Remark 2} The last equation uplifts this to Taking the Function Space Thereover 
(over configuration space in Geometry or over phase space in the canonical approach to Physics.)
This involves feeding in the `characteristic' solution $u$ of the first equation into the last equation by eliminating $\nu$.  
This $u$ then ends up featuring as the `functional form' that the observables depend on $\biQ$ and $\biP$ via, 
\be 
\bttO = \bttO(u)  \m .
\ee 
This simple outcome reflects that, in the strong case, the last equation is just a trivial ODE.  
The function space in question needs to be at least once continuously differentiable.
So the observables are suitably-smooth but elsewise-arbitrary functions of Lie's invariants (literally in Geometry, or their phase space counterparts in Physics).  

\m 

\n{\bf Remark 3} Solving for such arbitrary functions is to be contrasted with obtaining a single function by prescribing a specific boundary condition. 
{\sl Not} prescribing such a boundary condition, on the one hand, amounts to implementing Taking a Function Space Thereover. 
On the other hand, its more general technical name is {\it free} alias {\it natural} \cite{CH1} characteristic problem.  
The Characteristic Problem formulation for a single linear (or quasilinear) flow PDE is a standard prescription \cite{CH1, John}. 

\m 

\n{\bf Remark 4} At the geometrical level, our procedure is, given a constraint subalgebra $\scC_{\sfA}$, 
the observables equation $\mbox{\bf[} \scC_{\sfA} \mbox{\bf ,} \, \bttO \mbox{\bf ]}$ first determining a characteristic surface 
\be 
\chi = \chi(\biQ, \biP)                                                                                 \m .  
\ee
This follows from solving all but the last equation in the equivalent flow ODE system.  
Secondly, the last equation in the flow ODE system is a trivial ODE solved by any suitably-smooth function thereover. 
This gives the observables algebra as 
\be 
\CanObs(\lFrs)  \es  {\cal C}^{\infty}(\chi)  
             \es  \{ \m \mbox{ phase space functions whose restrictions to } \m \chi  \mma  \mf|_{\chi} \mma \mbox{ are smooth} \}  \m .  
\ee 
\n{\bf Remark 5} ${\cal C}^k$        for some fixed $k \geq 1$ could be used instead, 
                                      or some (perhaps weighted) Sobolev space \cite{AMP}; 
we adhere to     ${\cal C}^{\infty}$ for simplicity.

\subsection{$N$-point geometrical = purely configurational physical observables}

\n This subproblem has been covered in e.g.\ \cite{Lie80, Lie, G63, PE-1}. 

\m 

\n{\bf Definition 1} {\it classical geometrical observables} are 
\be 
\bttO(\biQ)                                               \m . 
\label{o-p}
\ee 
\n{\bf Structure 6} In the purely-geometrical setting, the a priori free functions $\bttO$ are subject to 
\be 
\mbox{\bf[} \sbcS \mbox{\bf,} \, \bttO \mbox{\bf]}  \es  0             \m ,
\label{SP}
\ee 
Here, $\sbcS$ is the {\it sum-over-N-points} \cite{G63, PE-1} $\u{q}^I$, $I = 1$ to $N$ of each particular generator, with components 
\be 
\scS_G   \:=  \sum_{I = 1}^N  {G_G}^b(q^{c I}) \frac{\pa \, }{\pa \,  q^{b I}}  \m .  
\label{S-Def}
\ee
\n{\bf Remark 1} As detailed in \cite{DO-1}, this is the pure-geometry analogue of $\sbcC$.  

\m 

\n{\bf Remark 2} The Lie bracket equation (\ref{SP}) can furthermore be written out as an explicit PDE system. 
It should by now be clear that this PDE is moreover a subcase of that for canonical observables in Theoretical Physics,  
\be 
\frac{\pa \,  \sbcC}{\pa \,  \biP} \frac{\pa \,  \bttO}{\pa \,  \biQ}  \es  0  \m ,
\ee
\be 
\mbox{with } \m \frac{\pa \,  \sbcC}{\pa \,  \biP} \m \mbox{ treated as knowns} \m .  
\ee 
In particular, $N$-point geometrical observables coincide with the purely configuration space restriction of physical observables \cite{DO-1}; 
see Sec \ref{Solns} for examples.

\subsection{Pure-momentum physical observables}

We furthermore consider the notion of {\it pure-momentum observables} 
\be 
\bttO(\biP)  \m . 
\label{o-p-2}
\ee 
These solve the {\it pure-momentum observables PDE system}  
\be 
\frac{\pa \,  \sbcC}{\pa \,  \biQ} \frac{\pa \,  \bttO}{\pa \,  \biP}  \es 0      \m ,
\ee
\be 
\mbox{for } \m  \frac{\pa \,  \sbcC}{\pa \,  \biQ} \m \mbox{ treated as knowns}  \m .
\ee 
\n{\bf Remark 1} Configuration and momentum observables each readily represent a {\it restriction} of functions over                           $\Phase$, 
                                                                                                   to      just over                           $\FrQ$, 
																							   and to           just over the space of momenta $\Mom$  respectively.   
These are, more specifically, {\it polarization restrictions} \cite{WoodhouseBook} since they precisely halve the number of variables. 
This applies at least in the quadratic theories we consider in the current Series.
These are by far the simplest and most standard form for bosonic theories in Physics.

\subsection{Various notions of genericity}

\n{\bf PDE genericity} From a PDE point of view, systems are more generic than single equations. 

\m 

\n{\bf Finite-theory geometrical genericity} However, in fixed-background finite theories of Geometry (or Physics), 
it is geometrically generic to have no (generalized) Killing vectors, and thus 0 or 1 observables PDEs. 

\m 

\n{\bf Remark 1} This is 0 in pure geometry and in temporally-absolute finite physics, 
                    to 1 in temporally-relational finite physics: commutation with $\chronos$.

\m

\n{\bf Remark 2} For Finite Theories, PDE system genericity is in general obscured by geometrical genericity. 

\m  

\n{\bf Remark 3} For Finite Theories, moreover, having 1 Killing vector is of secondary genericity between having no, and multiple, Killing vectors.  

\m 

\n{\bf Remark 4} From this point of view, unconstrained observables are most generic (no observables PDEs at all), 
                                          single observables PDEs are next most generic, 
										  and the more involved case of multiple observables equations is only tertiary in significance.

\subsection{Nontrivial system case: determinedness and integrability}

{\bf Remark 1} For nontrivial systems, multiple sequential uses of the Flow Method may apply.  
What needs to be checked first is determinedness \cite{CH2}, and, if over-determinedness occurs, integrability.  

\m 

\n{\bf Remark 2} In Geometry, we have $g := \mbox{dim}(\lFrg)$ constraints, and thus $g$ observables equations. 
The observables carry an index $\fO$ that has no a priori dependence on $\lFrg$.  
Thus, a priori, any of under-, well- or over-determinedness can occur (see also XIV.7).  
This conclusion transcends to the canonical approach to Physics as well. 
Here Temporal Relationalism and/or Constraint Closure can contribute further first-class constraints. 
$g$ is thus replaced by a more general count $f := \mbox{dim}(\FrF)$ of functionally-independent first-class constraints.   

\m 

\n On the one hand, generalized Killing equations' integrability conditions are not met generically (\cite{Yano55} or XIV.7).
This signifies that there are only any proper generalized Killing vectors at all in a zero-measure subset of $\langle \, \bFrM, \, \bsigma \, \rangle$. 
This corresponds to the generic manifold admitting no (generalized) symmetries.

\m 

\n On the other hand, preserved equations moreover always succeed in meeting integrability, by the following Theorem. 

\m 

\n{\bf Theorem 1} Classical canonical observables equations are integrable.

\m 

\n{\bf Remark 3} Consequently, classical observables {\sl always} exist (subject to the following caveats).  

\m 

\n{\bf Caveat 1} The current Article, and series, consider only {\it local} existence. 

\m 

\n{\bf Caveat 2} Sufficiently large point number $N$ is required in the case of finite point-particle theories.   
This is clear from the examples in \cite{PE-1, PE-2-3, DO-1}, 
and corresponds to zero-dimensional reduced spaces having no coordinates left to support thereover any functions of coordinates.] 

\m 

\n{\bf Remark 4} This Theorem is proven in \cite{DO-1}, resting on the following vaguely modern version of Frobenius' Theorem. 

\m 

\n{\bf Theorem 2 (A version of Frobenius' Theorem at the level of differentiable manifolds} \cite{AMP, Lee2}.  
A collection $\bFrW$ of subspaces of a tangent space possesses integral submanifolds iff\footnote{$\mbox{|[} \m \mbox{\bf ,} \, \m \mbox{\bf ]|}$ 
is here a general Lie bracket.} 
\be 
\forall \m \m \u{X} \, , \, \, \u{Y} \m \in \m \bFrW \mma \mbox{\bf |[} \u{X}\mbox{\bf ,} \, \u{Y} \mbox{\bf ]|} \m \in \m \bFrW   \m .
\ee

\subsection{Sequence of free characteristic problems for the strong observables system}

\n We now have a more extensive ODE system of the form (\ref{x-dot}, \ref{phi-dot}).

\m 

\n{\bf Remark 1} Regardless of the single-equation to system distinction, 
corresponding observables ODEs are moreover autonomous (none of the functions therein depend on $\nu$).  

\m 

\n{\bf Remark 2} We now have a first block rather than a first equation. 

\m 

\n{\bf Remark 3} While the system version is not a standard prescription; study of strong observables PDE systems gets past this 
by of our integrability guarantee.   

\m 

\n{\bf Sequential Approach}. 
Suppose we have two equations. 
Solve one for its characteristics $u_1$, say.   
Then substitute $\siQ = \siQ(u_1)$ into the second equation 
to find which functional restrictions on the first solution's characteristics the second equation enforces. 
This procedure can moreover be applied inductively. 
It is independent of the choice of ordering in which the restrictions are applied by the nature of restrictions corresponding to geometrical intersections.  

\m 

\n The Free Characteristic Problem posed above moreover leads to consideration of intersections of characteristic surfaces, 
which can moreover be conceived of in terms of restriction maps. 

\m 

\n{\bf Theorem 3} Suppose  
\be 
\bttV \m \mbox{ such that } \m \mbox{\bf [} \scC_{\sfV} \mbox{\bf ,} \, \bttV \mbox{\bf ]} = 0 \m \mbox{ forms characteristic surface } \m \chi_{\sttV}  
\ee 
and 
\be 
\bttW \m \mbox{ such that } \m \mbox{\bf [} \scC_{\sfW} \mbox{\bf ,} \, \bttW \mbox{\bf ]} = 0 \m \mbox{ forms characteristic surface } \m \chi_{\sttW}  \m 
\ee 
for constraint subalgebraic structures $\scC_{\sfV}$ and $\scC_{\sfW}$.  
Then 
$$ 
\bttO \m \mbox{ such that } \m \mbox{\bf [} \scC_{\tiV \cup \tiW} \mbox{\bf ,} \, \bttO \mbox{\bf ]} = 0 \m  
\mbox{ forms the characteristic surface of Fig \ref{Char-VIII}} \m .  
$$
%
{            \begin{figure}[!ht]
\centering
\includegraphics[width=0.4\textwidth]{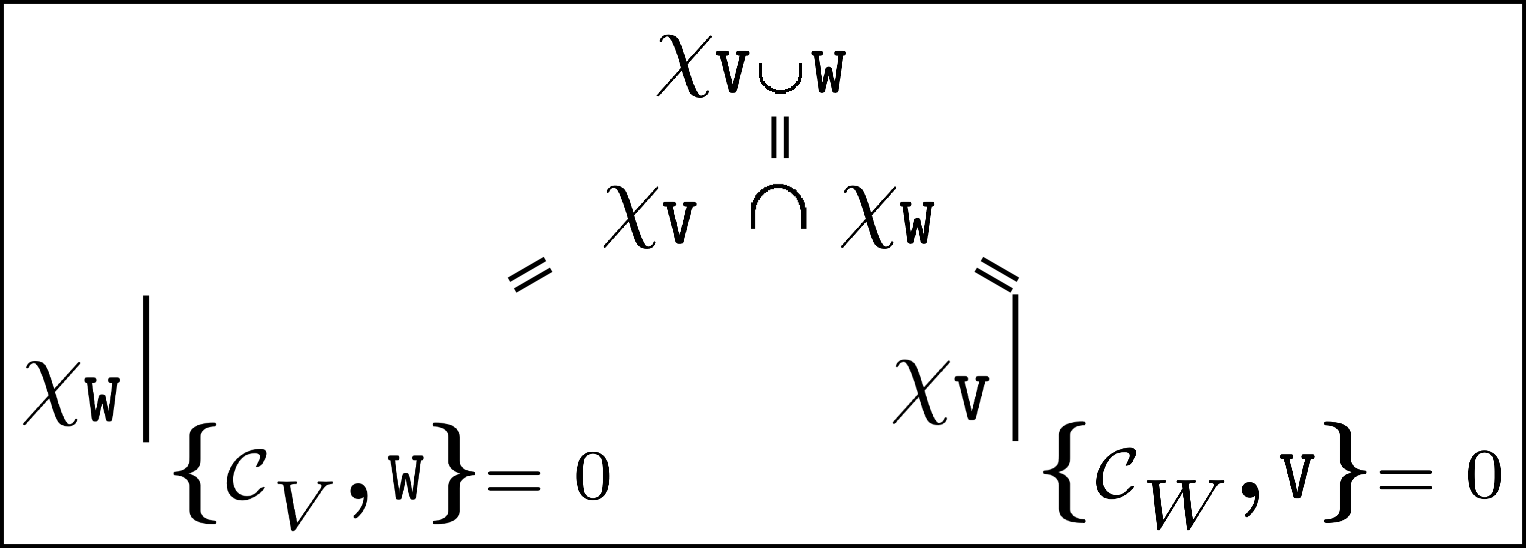} 
\caption[Text der im Bilderverzeichnis auftaucht]{    \footnotesize{Characteristic surface resulting from commutation with two constraint subalgebraic structures.} }
\label{Char-VIII} \end{figure}      }

\m 

\n{\bf Remark 4} This approach extends inductively to a finite number of equations in our flow ODE system. 

\m 

\n{\bf Remark 5} See e.g.\ \cite{PE-1, PE-2-3, DO-1} for examples of its use. 

\m 

\n{\bf Remark 6} The integrated form of the first $m$ equations is used to eliminate $t$, 
with the other $m - 1$ providing a basis of characteristic coordinates $u^{\bar{a}}$ arising as constants of integration. 
In the geometrical case, this can still be considered to be Lie's Method of Geometric Invariants. 
After all, essentially all the most familiar geometries involve more than one independent condition on their integral invariants. 

\m 

\n{\bf Remark 7} To elevate this to a determination of the system's observables, 
we then substitute these characteristic coordinates into the last equation. 
We thus obtain the general -- and thus free alias natural problem-solving -- characteristic solution.  

\m 

\n{\bf Remark 8} Our last equation remains a trivial ODE. 
It is thus solved by an suitably-smooth but elsewise arbitrary function of these characteristic coordinates, $\biu$ with components $u^{\bar{a}}$ 
\be 
\bttO = \bttO(\biu)
\ee   
\n{\bf Remark 9} The current Article (and Series) just considers a local rather than global treatment of observables equations.

\section{Weak observables PDEs}\label{W-PDE}

{\bf Structure 7} For weak observables, the brackets equation (\ref{W-Obs-Def}) gives 
\be 
\uc{\uo{\uc{\biW}}}  \, \uc{\sbcC}  \es  \mbox{\bf \{} \sbcC \mbox{\bf ,} \, \bttO \mbox{\bf \}}  
                                    \es  \frac{\pa \,  \sbcC}{\pa \,  \bQ}\frac{\pa \,  \bttO}{\pa \,  \bP} - \frac{\pa \,  \sbcC}{\pa \,  \bP} \frac{\pa \,  \bttO}{\pa \,  \bQ}  \m .  
\ee
We can again 
\be 
\mbox{take } \m  \frac{\pa \,  \sbcC}{\pa \,  \biQ} \m \mbox{ and } \m \frac{\pa \,  \sbcC}{\pa \,  \biP} \m \mbox{ to be knowns} \m , 
\ee 
now alongside $\sbcC$ being a known as well and $\biW$ taking some prescribed value. 
This leaves us with an inhomogeneous-linear first-order PDE system.  

\m 

\n{\bf Remark 1} The pure-geometry case cannot however support any properly weak observables. 
This is because we are in a context in which constraints have to depend on momenta
\be 
\sbcC = f(\biQ, \, \biP) \m \mbox{ \sl with specific $\biP$ dependence } \m .   
\ee
A fortiori, first-class linear constraints refer specifically to being linear in their momenta $\biP$, 
By this, the weak configurational observables equation would have a momentum-dependent inhomogeneous term right-hand side. 
But this is inconsistent with admitting a solution with purely $\biQ$-dependent right-hand-side.

\subsection{Single weak observables equation}

\n The weak observables PDE consists of a single inhomogeneous-linear equation. 
Its corresponding ODE system is now of the form 
\be 
\dot{x} = a(x, \phi)  \m , 
\ee 
\be 
\dot{\phi} = b(x, \phi)  =  W \, \scC_1             \m .
\ee
\n{\bf Remark 1} The first block is the same as before. 
Lie's integral invariants (in Geometry or their canonical Physics generalizations) thus still enter our expressions for observables. 

\m 

\n{\bf Remark 2} The inhomogeneous term in the last equation, however, means that one has further particular-integral work to do in this weak case.

\subsection{Nontrivial weak observables system}

The corresponding ODE system now has inhomogeneous term $b_{\sfO} = {W^{\sfB}}_{\sfA \sfO} \, \scC_{\sfB}$.  

\m 

\n{\bf Remark 1} Determinedness and integrability considerations carry over. 
So does sequential use of free characteristic problems on the first block. 

\m 

\n{\bf Remark 2} Solving the last equation in the system is then conceptually the same as in the previous subsection.  

\m 

\n{\bf Remark 3} We leave a systematic Green's function approach to weak observables equations for another occasion. 

\m 

\n{\bf Remark 4} If we reduce all constraints out, the reduced formulation has strong $\widetilde{\bttU}$ being all the observables there can be. 

\m 

\n{\bf Remark 5} If we reduce all constraints bar $\chronos$ out, the following applies since finite theory's Chronos is a single constraint equation. 

\m 

\n{\bf Corollary 2} Suppose there is only one constraint. 
Then there is only one independent proper weak observable, and the proper weak observables algebra is abelian. 

\m 

\n{\u{Proof}} The PI is now a single integral up to a multiplicative constant. 

\m  

\n Any finite-algebra generator commutes with itself.  $\Box$ 

\m 

\n{\bf Remark 6} Let us check how this comes to be consistent with Corollary 1.

\m 

\n The first part of the first factor of (\ref{factors})'s antisymmetry is not supported in nontrivially-weak observables space of dimension 1. 
(\ref{factors}) thus collapses to the furtherly factorizable form  
\be 
O^{\sw} W_{\sfO} \, \scC = 0  \m .  
\ee 
$\scC = 0$ is disallowed as $\scC$ is a nontrivial constraint, whereas $W_{\sfO} := {W^1}_{1\sfO} = 0$ is disallowed as $O^{\sw}$ is nontrivially-weak.
The there is no space for the zero double-trace perpendicularity as each trace is now over a 1-$d$ index and the 1 \time 1 matrix 1 is manifestly nondegenerate. 
Thus $O^{\sw} = 0$ is the only surviving possibility, coinciding with the above deduction that the nontrivially-weak observables algebra is abelian.

\section{Expression in Terms of Observables}\label{EiToO}

\n We here continue to develop Sec III.3. 
First recall that, by Lemma III.3, the $\observables$ themselves form a closed algebraic structure.  
The amount of observables, is, moreover, very large due to the following Lemma. 

\m 

\n{\bf Lemma 3} Functions of observables are themselves observables.  

\m 

\n{\u{Proof}} 
\be 
\mbox{\bf \{} \sbcC \mbox{\bf ,} \, F(\bttO) \mbox{\bf \}}        \es      \mbox{\bf \{} \sbcC \mbox{\bf ,} \, \bttO \mbox{\bf \}}  \,  \frac{\pa \,  F}{\pa \,  \bttO} 
                                                            \m \approx \m                         0                  \times             \frac{\pa \,  F}{\pa \,  \bttO}   
                                                                  \es                             0
															\m . \m \m  \Box
\ee 
\n{\bf Remark 1} Thus one is not looking for individual solutions of the observables PDEs, but a fortiori for whole algebras of solutions. 
Namely, it is closed among themselves and large enough to span all of a physical theory's mathematical content.  
I.e.\ enough to span $\Phase$, with each functionally independent of the others. 
This renders useful the concept mentioned in Article III of finding `basis observables'. 

\m 

\n{\bf Remark 2} Only a smaller number of suitably-chosen observables moreover suffice for practical use: expression in terms of observables. 

\m 

\n{\bf Remark 3} A common case is for $\mbox{dim(reduced $\Phase$)} = 2\{k - g\}$ basis observables (of type $\bttG$) to be required. 
Here $k := \mbox{dim}(\FrQ)$ and $g := \mbox{dim}(\lFrg)$ is the total number of constraints involved, which are all gauge constraints. 
See e.g.\ Sec \ref{BB} for examples of `basis observables'.

\section{Strategies for the Problem of Observables}\label{Strat}

\n The `bottom' alias `zero' and `top' alias `unit' strategies are as follows. 

\m 

\n{\bf Strategy 4.0)} {\sl Use Unconstrained Observables}, $\unres$ \cite{RovelliBook, Ditt}, entailing no commutation conditions at all.

\m 

\n{\bf Strategy 4.1)} {\sl Insist on Constructing Dirac Observables}, $\Dirac$.

\m

\n{\bf Remark 1} Strategy 4.1) has the conceptual and physical advantage of employing all the information 
in the final algebraic structure of all-first-class constraints of the theory, $\FrF$. 
It has the practical disadvantage that finding any Dirac observables -- much less a basis set for each theory in question -- can be a hard mathematical venture, 
especially for Gravitational Theories \cite{DiracObs, Bergmann61, BK72, HTBook, K92, I93, K93, RovelliBook, Ditt, G-M-H, ThiemannBook, PSS10, AObs, DHKN15, ABook}. 

\m 

\n{\bf Remark 2} Strategy 4.2) is diametrically opposite in each of the above regards.  
It can moreover be used as first stepping stone toward the former.   

\m

\n{\bf Remark 3} Strategy 4.1) moreover amounts to concurrently addressing the unsplit totality of constraints (Constraint Closure facet) 
                                                                                                           and  Taking Function Spaces Thereover.  
As a four-aspect venture (Fig \ref{FST-EitoB-Interference}), it is unsurprisingly harder than Use Unconstrained Observables, which is single-aspect.  

\m 

\n{\bf Strategy 4.K)} {\sl Find \K Observables} $\Kuchar$.   
This can entail treating $\Quad$ distinctly from the $\bFlin$, some motivations for which were covered in VII. 
A further pragmatic reason is that the $\Kuchar$ are simpler to find than the $\Dirac$.

\m 

\n{\bf Strategy 4.G)} {\sl Find} $\lFrg${\sl -observables}, $\gauge$.   
If one looks more closely, some of these motivations are actually tied to the $\gauge$ in cases in which these and the $\Kuchar$ are distinct.   
For instance, it is more generally $\gauge$ -- rather than the $\Kuchar$ -- 
which arise from the $\lFrg$-act, $\lFrg$-all construction in cases in which the candidate $\bShuffle$ is confirmed as a $\bGauge$.
Also, theories having either trivial Configurational Relationalism -- or Best Matching resolved -- 
have as a ready consequence a known full set of classical $\gauge$.
We take this on board by pointing to this further distinct strategy.
This takes into account the triple combination of Configurational Relationalism, 
                                                  Constraint Closure, and 
												  Assignmant of Observables aspects.
Complementarily, finding the $\gauge$ (or $\Kuchar$) represents a timeless pursuit 
due to the absence of $\Chronos$ (or underlying Temporal Relationalism) from the workings in question.  

\m

\n{\bf Strategy 4.C)} {\sl Find Chronos Observables}, $\chronos$. 
This takes into account the triple aspect combination of Temporal Relationalism, Constraint Closure and Taking Function Spaces Thereover.  
In theories with nontrivial $\lFrg$ or some further first-class constraints, this is to be viewed as a stepping stone. 
It is available if $\Chronos$ indeed constitutes a constraint subalgebraic structure.  

\m 
  
\n{\bf Remark 4} (Non)universality arguments are pertinent at this point.
Using $\unres$ or $\Dirac$ is always in principle possible.
The first of these follows from no restrictions being imposed.
The second follows from how any theory's full set of constraints can in principle be cast as a closed algebraic structure of first-class constraints. 
This is by use of the Dirac bracket, or the effective method, so as to remove any second-class constraints.

\m 

\n{\bf Strategy 4.A)} We finally introduce an additional universal strategy based on {\sl using some kind of A-Observables}, 
$\bttA_{\sfx}$, {\sl that a theory happens to possess}.  
This corresponds to the closed subalgebraic structures of constraints which are realized by that theory.  
It is the general `middling' replacement for considering $\Kuchar$, $\Gaugeo$ or $\Chronos$  observables.
(None of these notions are universal over all physical theories.)   
While the ultimate aim is to reach the top of the lattice, it is often practically attainable to, firstly, land somewhere in the middle of the lattice.
Secondly, to work one's way up by solving further DEs (or, geometrically, by further restricting constraint surfaces).  

\m 

\n A second source of strategic diversity is as follows.  

\m 

\n{\bf Strategy 4$^{\prime}$.0)}  The {\it unreduced approach}: working on the unreduced $\Phase(\lFrs)$ with all the constraints.

\m

\n{\bf Strategy 4$^{\prime}$.1)} The {\it true space approach} -- working on $\TruePhase(\lFrs) = \Phase(\lFrs)/\FrF$ -- is much harder, 
due to quotienting out $\Chronos(\lFrs)$ being harder and having to be done potential by potential.
\n If True is known, classical observables are trivial.  

\m 

\n{\bf Strategy 4$^{\prime}$.G)} The {\it reduced approach}: working on $\w{\Phase}(\lFrs) = \Phase(\lFrs)/\lFrg$ with just $\Chronos$. 

\m 

\n{\bf Strategy 4$^{\prime}$.P)}  The {\it partly-reduced approach}: working on $\Phase(\lFrs, \FrH) = \Phase(\lFrs)/\FrH$ for some $id < \FrH < \lFrg$ 

\m 

\n{\bf Remark 5} The above primed family of strategies are moreover found to all coincide in output in the case of strong observables \cite{DO-1}. 
E.g.\ both unreduced and reduced approaches end up with 
\be
\CanGaugeObs(\lFrs)  \es {\cal C}^{\infty}(\w{\Phase}(\lFrs)) 
\ee
or, in the purely geometrical case,  
\be
\GeomGaugeObs(\lFrs)  \es {\cal C}^{\infty}(\w{\FrQ}(\lFrs)) \m .  
\ee

\section{Explicit solutions for observables and observables algebras}\label{Solns}

\subsection{Unreduced examples}

\n Unreduced observables form the function space 
\be 
\CanUnresObs(\lFrs)  \es  {\cal C}^{\infty}(\Phase(\lFrs))                         
\label{U-Phase}
\ee 
of suitably-smooth functions over our system's phase space. 

\m 

\n Unreduced configurational observables form the function space 
\be 
\GeomUnresObs(\lFrs)  \es  {\cal C}^{\infty}(\FrQ(\lFrs))                         
\label{U-Q}
\ee 
of suitably-smooth functions over our system's configuration space.

\m 

\n Unreduced pure-momentum observables form the function space 
\be 
\MomUnresObs(\lFrs)  \es  {\cal C}^{\infty}(\Mom(\lFrs))                         
\label{U-P}
\ee 
of suitably-smooth functions over our system's {\it momentum space} 
\be 
\Mom(\lFrs)  
\ee 
that consists of the totality of values taken by the given model $\lFrs$'s momenta $\biP$. 

\m 

\n{\bf Example 1} For Mechanics on $\mathbb{R}^d$, the configuration space is the constellation space (I.5-6), $\FrQ(d, N) = \mathbb{R}^{d \, N}$. 
This being a flat space, the corresponding {\it constellation phase space} is 
\be 
\Phase(N, d) := \mathbb{R}^{2 \, N \, d}   \m .
\ee 
The unrestricted observables 
\be
\bttU  \es  \bttU(\biq, \, \bip)            
\label{U(p,q)}
\ee
here form 
\be 
\CanUnresObs(d, N)  \:=  {\cal C}^{\infty}(\Phase(\bFrM, N))  
                    \es  {\cal C}^{\infty}(\mathbb{R}^{2 \, N \, d})  \m . 
\ee
The unrestricted configurational observables $\bttU(\biq)$, alias absolute space geometry's $N$-point invariants, form 
\be 
\GeomUnresObs(d, N)  \:=  {\cal C}^{\infty}(\FrQ(d, N))  
                     \es  {\cal C}^{\infty}(\mathbb{R}^{d \, N})      \m .    
\ee 
Finally the unrestricted pure-momentum observables $\bttu(\bip)$ form 
\be 
\MomUnresObs(d, N)  \:=  {\cal C}^{\infty}(\FrP(d, N))  
                    \es  {\cal C}^{\infty}(\mathbb{R}^{N \, d})       \m .  
\ee
These results readily generalize to RPMs over manifolds other than $\mathbb{R}^n$.

\subsection{Nontrivial geometrical examples of Kucha\v{r}, gauge- and A-observables}	

\n{\bf Simplification 1} Suppose the constraints being taken into consideration depend at most linearly on the momenta.
This holds within $Aff(d)$ and its lattice of subgroups, covering all but the last example in the current subsection.  
Then in the observables equations, 
\be 
\mbox{the cofactor of } \m  \frac{ \pa \,  \bttG }{ \pa \,  \biQ } \m 
\mbox{ -- i.e.\ }       \m  \frac{ \pa \,  \sbcC }{ \pa \,  \biP } \m \mbox{ -- is independent of } \m  \biP \m .  
\ee 
\n{\bf Simplification 2} Restrict attention to purely configurational such observables $\Kuchar(\biQ) = \Gaugeo(\biQ)$.
This gives the particularly simple PDE [following on from (VII.22)]
\beq
{{\cal F}^{\sfA}}_{\sfN}(\biQ \mbox{ alone}) \frac{\pa \,  \Gaugeo}{\pa \,  Q^{\sfA}}  \es  0  \m ;  
\label{Cl-K-B-Eq}
\eeq
%
%
we drop the $\sbiQ$ suffix when considering pure geometry.)

\m 

\n{\bf Simplification 3} A few of the below examples have but a single observables equation (i.e.\ the $\fN$-index takes a single value). 
This is then amenable to the standard Flow Method. 

\m 

\n{\bf Simplification 4} Within $Aff(d)$'s lattice of subgroups, passage to the centre of mass frame is available to take translations out of contention.  
Upon doing this, mass-weighted relative Jacobi coordinates \cite{Marchal, II} 
furthermore furnish a widespread simplification of one's remaining equations \cite{FileR, S-I}. 
This determines ab-initio-translationless and translation-reduced actions, constraints, Hamiltonians and observables,  
as being of the same form but with one object less: the so-called {\it Jacobi map} \cite{S-I}.   

\m 

\n{\bf Notation 1} We additionally provided a compact notation for the outcome of solving the preserved equations for $Sim(d)$ 
and its subgroups \cite{S-I, Minimal-N} in Sec III.3. 
This is visible within row 1 of Fig \ref{VIII-Fig}, which furthermore extends this notation to $Aff(d)$ and its subgroups \cite{AMech, PE-2-3}.  
%
{            \begin{figure}[!ht]
\centering
\includegraphics[width=1.0\textwidth]{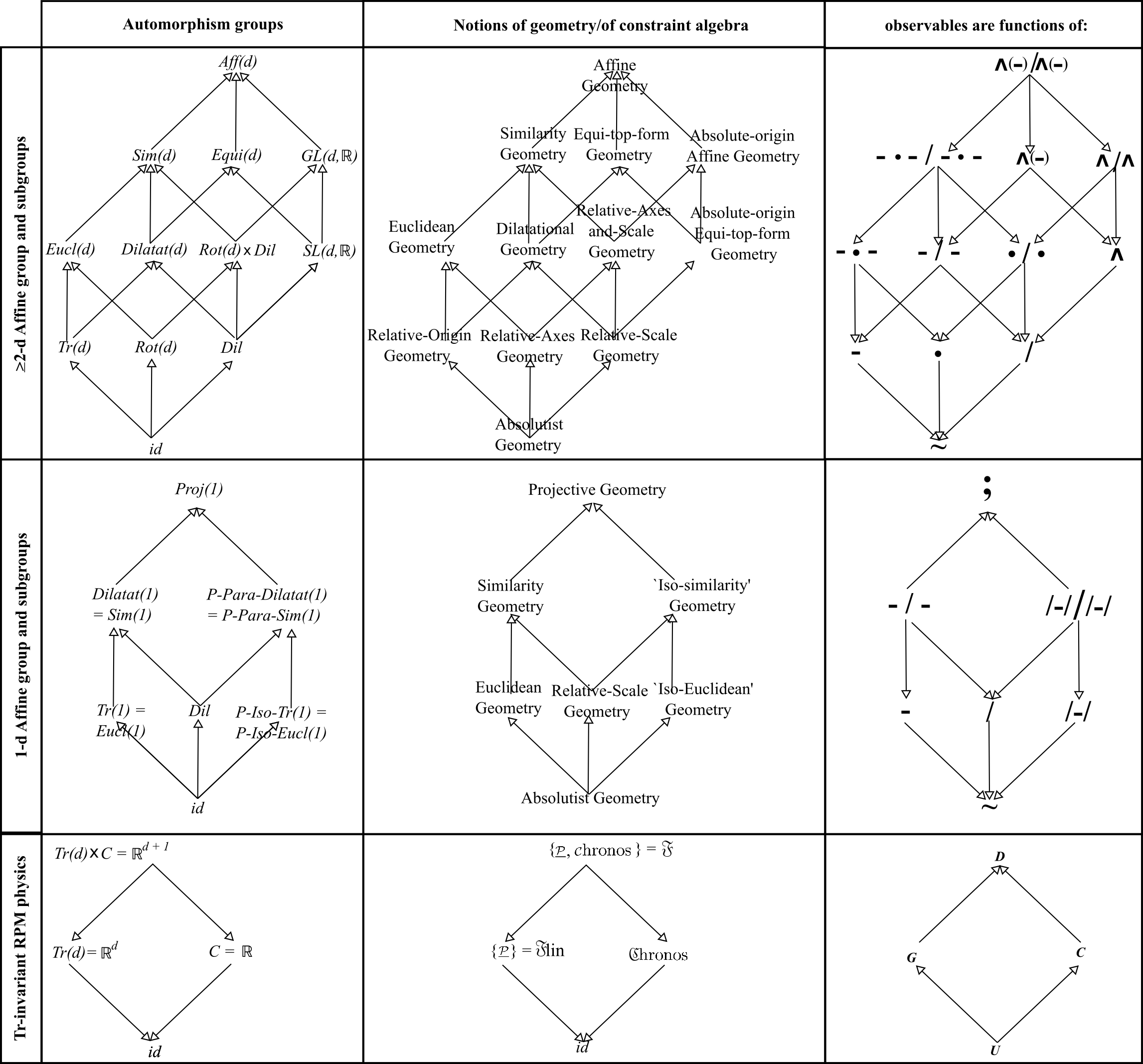}
\caption[Text der im Bilderverzeichnis auftaucht]{        \footnotesize{a) As a reminder of Article III's notation, the minus stands for difference, 
                                                                                                                    the backslash    for ratio,          and  
																												    the dot          for scalar product.  
The new wedge symbol stands for the `top form' supported by the dimension in question.  
So e.g.\ area's cross product in 2-$d$ or volume's scalar triple product in 3-$d$. 
The notation for this symbol is slightly different than for the others, since it is an $n$-ary rather than binary operation.   

\m 

\n b) The semicolon here denotes cross ratio. 
See \cite{PE-2-3} for what any notions of geometry mentioned in the first two rows mean.  
I subsequently found that Guggenheimer \cite{G63}, 
and then that previously Lie himself \cite{Lie80} already derived the top result here -- the 1-$d$ projective group -- by a flow PDE method. 
So (for now at least), the top case of Example 8 is to be attributed to Lie: an early success of his Integral Approach to Geometrical Invariants.

\m 

\n c) is derived in detal in \cite{DO-1}.}}
\label{VIII-Fig} \end{figure}          }

\m 

\n{\bf Example 1)} For translation-invariant geometry -- $\lFrg = Tr(d) $ -- the observables PDE is 
\beq
\sumIN \frac{\pa \,  \Gaugeo}{\pa \,  \u{q}^I}                                                                        \es  0  \m .
\label{Ku-P}
\eeq 
The solutions of this are, immediately, the relative interparticle separation vectors. 
These can be reformulated as linear combinations thereof, 
among which the relative Jacobi coordinates $\brho$ turn out to be particularly convenient by the Jacobi map.    
Our solutions form 
\be 
\GeomGaugeObs(d, N; Tr(d))  \es  {\cal C}^{\infty}(\mathbb{R}^{n \, d}) \m : 
\ee 
the smooth functions over relative space.  

\m 

\n{\bf Example 2} For scale-invariant geometry -- $\lFrg =  Dil$ -- the observables PDE is  
\beq
\sumIN \underline{q}\mbox{}^I \cdot \frac{\pa \,  \Gaugeo}{\pa \,  \underline{q}\mbox{}^I}   \es   0  \m . 
\label{Ku-D}
\eeq
This is an Euler homogeneity equation of degree zero.  
Its solutions are therefore ratios of components of configurations.
These form 
\be 
\GeomGaugeObs(d, N; Dil)  \es  {\cal C}^{\infty}(\mathbb{S}^{N \, d - 1})  \m :
\ee 
the smooth functions over ratio space. 

\m 

\n{\bf Example 3} For rotationally-invariant geometry -- $\lFrg = Rot(d)$ -- the observables PDE is 
\beq
\sumIN \u{q}^I \cr \frac{\pa \,  \Gaugeo}{\pa \,  \u{q}^I}   \es  0  \m . 
\label{Ku-L}
\eeq
This is solved by the dot products $\u{q}^I \cdot \u{q}^J$. 
Norms and angles are moreover particular cases of functionals of the above, which are an allowed extension by Lemma 3.
In 2-$d$, these form 
\be 
\GeomGaugeObs(2, N; Rot(2))  \es  {\cal C}^{\infty}(C(\mathbb{CP}^{n})) \m , 
\ee 
where $C(\FrM)$ denotes topological and geometrical cone over $\FrM$ and $n := N - 1$.  

\m 

\n{\bf Example 4} In Euclidean geometry --$\lFrg = Eucl(d) = Tr(d) \rtimes Rot(d)$ -- the observables PDEs are both (\ref{Ku-P}) and (\ref{Ku-L}). 
Sequential use of the Flow Method gives that the solutions are dots of differences of position vectors.
This can be worked into the form of dots of relative Jacobi vectors.
These are additionally Euclidean RPM's \cite{BB82, FileR} configurational \K = gauge observables \cite{AObs2}.
In 2-$d$, they form 
\be 
\GeomGaugeObs(2, N; Eucl(2))  \es  {\cal C}^{\infty}(C(\mathbb{CP}^{n - 1}))  \m :
\ee
the smooth functions over relational space. 

\m  

\n{\bf Example 5} In rotational-and-dilational geometry -- $\lFrg = Rot(d) \times Dil$ -- the observables PDEs are both (\ref{Ku-L}) and (\ref{Ku-D}). 
The geometrical $\Gaugeo$ are thus ratios of dots. 
In 2-$d$, these form 
\be 
\GeomGaugeObs(2, N; Rot(2) \times Dil)  \es  {\cal C}^{\infty}(\mathbb{CP}^{n}) \m .  
\ee 
\n{\bf Example 6} In dilatational geometry -- $\lFrg = Dilatat(d) := Tr(d) \rtimes Dil$ -- the observables PDEs are both (\ref{Ku-P}) and (\ref{Ku-D}). 
The geometrical $\Gaugeo$ are thus ratios of differences. 
These form 
\be 
\GeomGaugeObs(d, N; Dilatat(d))  \es  {\cal C}^{\infty}(\mathbb{S}^{n \, d - 1})  \m :  
\ee 
the smooth functions over preshape space \cite{Kendall}. 

\m 

\n{\bf Example 7} In similarity geometry -- $\lFrg = Sim(d) = Tr(d) \rtimes \left( Rot(d) \times Dil \right)$ --  
the observables PDEs are all three of (\ref{Ku-P}), (\ref{Ku-L}) and (\ref{Ku-D}). 
The geometrical $\Gaugeo$ are thus ratios of dots of differences. 
These are additionally similarity RPM's \cite{FileR} configurational \K observables $\Kuchar$ \cite{AObs2}.
In 2-$d$, these form 
\be 
\GeomGaugeObs(2, N; Sim(2))  \es  {\cal C}^{\infty}(\mathbb{CP}^{n - 1}) \m :  
\ee 
the smooth functions over shape space \cite{Kendall}. 

\m 

\n{\bf Example 8} Row 1 of Fig \ref{VIII-Fig} identifies geometrical observables for Affine Geometry -- corresponding to $\lFrg = Aff(d)$ -- 
and for various of its further subgroups.    
We need dimension $d \geq 2$ for these groups to be nontrivially realized.
These additionally constitute configurational observables for affine RPM \cite{AMech}.  

\m 

\n Each of Examples 1 to 7 also constitute nontrivially-$\bttA$ examples (in the present context $\bttA$ that are not also $\Kuchar = \Gaugeo$) for affine geometry, 
as do all other entries in the third subfigure of Fig \ref{VIII-Fig}.a) 
Each of Examples 1 to 6 perform this function for similarity geometry as well. 
Two-thirds of Examples 1 to 3 perform this function for each of Examples 4 to 6. 
Flat Geometry thus already provides copious numbers of examples of nontrivially-$\bttA$ observables.

\m 

\n{\bf Example 9} Row 2 of Fig \ref{VIII-Fig} identifies geometrical observables for 1-$d$ Projective Geometry -- corresponding to $\lFrg = Proj(1)$ -- 
and various of its further subgroups.    
Passing to the centre of mass however ceases to function as a simplification for Projective Geometry (and Conformal Geometry).   
This is since translations are more intricately involved here than via Affine Geometry's semidirect product addendum.

\subsection{Geometrical examples of `basis observables'}\label{BB}  

We here provide Euclidean RPM $\Kuchar = \bttG$ observables examples of this. 

\m  

\n{\bf Example 1} 3 particles in 1-$d$ have the mass-weighted relative Jacobi separations $\rho_1, \rho_2$ as useful basis observables (in $\Kuchar = \bttG$ sense).
This extends to $N$ particles in 1-$d$ having as basis observables a given clustering's $\rho^i$, $i = 1$ to $n = N - 1$.  

\m 

\n{\bf Example 2} The 3 Hopf--Dragt coordinates of Sec V.5.5 are basis observables for the relational triangle \cite{FileR}. 

\m 

\n{\bf Example 3} The 8 `Gell--Mann' coordinates detailed in \cite{Quad-I} are basis observables for the relational quadrilateral. 

\m 

\n{\bf Motivation} Kinematical quantization uses a lot less classical observables than the totality of suitably smooth functions over $\Phase$.
Kinematical quantization uses, more specifically, a linear subspace thereof that the canonical group acts upon \cite{I84}.  
This linear subspace moreover has enough coordinates to locally characterize $\FrQ$, thus fitting within our looser conception of `basis beables'. 
Kinematical quantization's linear subspace quantities do moreover {\sl literally} form a basis for that linear space.
In this way, they constitute `more of a basis' for configuration space than the a priori concept of `basis observables' do.  
This linearity does not however in general extend to kinematical quantization's corresponding momentum observables (angular momentum suffices to see this).  
All of our examples above are useful for kinematical quantization.

\subsection{Physically nontrivial examples of strong $\Kuchar$, $\bttG$, $\bttA$ and $\bttC$ observables}

\n{\bf Example 1} Translationally-invariant RPM's pure-momentum observables are freely specifiable.  
However, since the total centre of mass position is meaningless in this problem, its momentum is meaningless as well, leaving us with 
\be  
\sbiG  \es  \sbiG \left( \u{p}_i - \u{p}_N \right) 
       \es  \sbiG \left( \u{\pi}_i \right)
%
%
\ee
for $\u{\pi}_i$ the conjugate momenta to $\u{\rho}^i$.
These form 
\be 
\MomGaugeObs(d, \, N; \, Tr)  \es  {\cal C}^{\infty}(\mathbb{R}^{n \, d})           \m ,  
\ee
i.e.\ the smooth functions over relative momentum space. 

\m 

\n The corresponding general observables are 
\be 
\sbiG  \es  \sbiG \left( \u{q}^i - \u{q}^N , \, \u{p}_i - \u{p}_N  \right)  
       \es  \sbiG \left( \u{\rho}^i , \, \u{\pi}_i                 \right)          \m .  
\label{Tr-Obs}
\ee    
These form 
\be 
\CanGaugeObs(d, \, N; \, Tr(d))  \es  {\cal C}^{\infty}(\mathbb{R}^{2 \, n \, d})   \m ,  
\label{Tr-Can-Gauge-Obs}
\ee 
i.e.\ the smooth functions over relative phase space. 

\m 

\n{\bf Example 2} Dilationally-invariant RPM's pure-momentum observables PDE is the 
\be 
\u{q}^I \leftrightarrow \u{p}^I
\label{p<->q}
\ee 
of the corresponding geometrical observables PDE system. 
This is thus solved by ratios of components of momenta.  
These form 
\be 
\MomGaugeObs(c, \, N;  Tr(d)) \es {\cal C}^{\infty}(\mathbb{S}^{N \, d - 1}) \m , 
\ee 
i.e.\ the smooth functions over momentum ratio space.

\m 

\n The corresponding general observables PDE system is 
\beq
\sumIN 
\left\{   \frac{\pa \,  \bttG}{\pa \,  \u{p}^I} \cdot \u{p}_I \m \m - \m \m \u{q}^I \cdot \frac{\pa \,  \bttG}{\pa \,  \u{q}^I}  \right\}  \es  0                 \m . 
\label{Ku-D-Full}
\eeq
This is also an Euler homogeneity equation of degree zero.  
Its solutions are therefore ratios of phase space coordinates.  
These now correspond not to sphere in phase space but to a quadric surface 
\be 
\sumIN \left(  ||\u{p}_I||^2 - ||\u{q}^I||^2 \right)  \es  const
\ee 
in phase space. 
The observables constitute the $\mathbb{C}^{\infty}$ functions over this phase ratio space. 

\m 

\n{\bf Example 3} Rotationally-invariant RPM's pure-momentum observables PDE system is also 
the (\ref{p<->q}) of the corresponding geometrical observables PDE system. 
This is thus is solved by suitably smooth functions of the dot product,  
\be 
\bttG  =  \bttG(\u{p}^I \cdot \u{p}^J)  \m . 
\ee 
In 2-$d$, these form 
\be 
\MomGaugeObs(d, N; Rot(2)) \es {\cal C}^{\infty}(C(\mathbb{CP}^{N})) \m .  
\ee    
The corresponding general observables PDE system is 
\be 
\sumIN 
\left\{   \frac{\pa \,  \bttG}{\pa \,  \u{p}^I} \cr \u{p}_I \m \m + \m \m \frac{\pa \, \bttG}{\pa \, \u{q}^I} \cr \u{q}^I  \right\}  \es  0  \m . 
\label{Ku-L-Full} 
\ee 
This is solved by suitably-smooth functions of 
\be 
\cdot_{\sbS}  \:=  \u{q}^I \cdot \u{p}^J + \u{p}^I \cdot \u{q}^J  \m : 
\ee  
phase space symmetrized dot products. 
These are the outcome of applying the product rule to $\u{q}^I \cdot \u{q}^J$. 

\m 

\n{\bf Example 4} Euclidean RPM combines the above translational and rotational equations. 
The Jacobi map applying, the Euclidean momentum observables solutions are suitably smooth functions  
\be 
\bttG = \bttG(\u{\pi}^i \cdot \u{\pi}^j) \m   
\ee
In 2-$d$, these form 
\be 
\MomGaugeObs(d, N; Eucl(2)) \es {\cal C}^{\infty}(C(\mathbb{CP}^{n})) \m :   
\ee   
the smooth functions over relational momentum space. 
The general observable solutions are suitably smooth functions 
\be 
\bttG = \bttG(\u{\rho}^i \cdot \u{\pi}^j + \u{\pi}^i \cdot \u{\rho}^j) = \bttG(-\cdot_{\sbS}-) \m , 
\ee 
i.e.\ relative phase space symmetrized dot products. 

\m 

\n{\bf Example 5} Similarity RPM combines all three of the above translational, rotational and dilational equations. 
So for instance, the momentum observables solutions are suitably smooth functions  
\be 
\bttG = \bttG(\u{\pi}^i \cdot \u{\pi}^j/\u{\pi}^k \cdot \u{\pi}^l)  \m ,  
\ee 
and the general observables are suitably smooth functions 
\be 
\bttG = \bttG(-\cdot_{\sbS}- / -\cdot_{\sbS}- )  \m . 
\ee 
\n{\bf Example 6} If the generators are quadratic (which we know from Article III to apply in the conformal and projective cases), 
then (\ref{p<->q}) symmetry among observables is broken. 

\m 

\n{\bf Example 7} Chronos observables $\bttC$ for the general $(N, d)$ Euclidean RPM solve 
\be 
\bip \cdot \frac{\pa \,  \bttC}{\pa \,  \,  \biq}  \m + \m   \frac{\pa \,  V}{\pa \,  \biq} \frac{\pa \, \bttC}{\pa \, \biq}\es  0  \m .  
\ee 
In the case of constant potential, this simplifies to 
\be 
\bip \cdot \frac{\pa \, \bttC}{\pa \, \biq}  \es  0
\ee 
which is solved by  
\be 
\sbiC  \es  \sbiC  \left(  q^{\Lambda} p_{N \, d}  -  p_{\Lambda} q^{N \, d} , \,{p_{\Lambda}}^2 - {p_{N \, d}}^2  \right)  \m . 
\ee
Treating the $N \, d$ component differently is an arbitrary choice; $\Lambda$ then runs over all the other values, $1$ to $N \, d - 1$.  
For RPMs with any Configurational Relationalism, these are a further species of $\bttA$ observable that is not a $\Kuchar$ or $\bttG$ observable.  

\m 

\n{\bf Remark 1} That $\ChronosObs$ (and $\DiracObs$) are potential-dependent, constitutes a massive complication at the computational level.

\subsection{Examples of strong nontrivially-Dirac observables}

\n{\bf Example 1} For RPMs involving whichever combination of translations and rotations, the additional PDE to obtain strong Dirac observables is 
\be 
\sumIN 
\left\{  \frac{ \pa \, V }{ \pa \,  \u{q}^I } \cdot \frac{ \pa \,  \Dirac }{ \pa \,  \u{p}_I }  \m - \m  
                                    \u{p}_I   \cdot \frac{ \pa \, \Dirac }{ \pa \,  \u{q}^I }  \right\}  \peq  0  \m .  
\ee
If dilations are involved as well, one needs to divide the kinetic term contribution by the total moment of inertia to have a ratio form.  

\m 

\n The above are moreover mathematically equivalent to the corresponding $(d, \, n)$ Chronos problems, so e.g.\ in the $Eucl(d)$ case 
\be 
\w{\sbiD}  \es  \w{\sbiD}\left(\rho^{\tau} \pi_N - \rho^N \pi_{\sigma} , \, {\pi_{\tau}}^2 - {\pi_N}^2\right)   \m ,  
\ee 
with the true space index $\tau$ running over 1 to $n \, d - 1$.

\m 

\n{\bf Example 2}  {\it Minisuperspace} (spatially homogeneous GR) only has a $\scH$, 
and a single finite constraint oversimplifies the diversity of notions of observables. 
Here the sole strong observables brackets equation is 
\be
\mbox{\bf \{} \,    \scH    \mbox{\bf ,} \, \Dirac    \, \mbox{\bf \}}   \es  0                              \m , 
\ee
giving the observables PDE 
\be 
\frac{\pa \,  \scH}{\pa \,  \biQ} \frac{\pa \,  \Dirac}{\pa \,  \biP}  \m - \m  \frac{\pa \,  \scH}{\pa \,  \biP}  \frac{\pa \,  \Dirac}{\pa \,  \biQ}  \es  0  
\ee
\be  
\mbox{for } \m  \frac{\pa \,  \scH}{\pa \,  \biQ}  \m \mbox{ and } \m  \frac{\pa \,  \scH}{\pa \,  \biP} \m \mbox{ knowns} \m .
\ee  
As per Article I, simple examples include $\biQ = \alpha, \phi$ or $\alpha, \beta_{\pm}$.  
Each minisuperspace's type of potential (one part fixed by GR, another part variable with the nature of appended matter physics).  

\m 

\n There being just one such equation gives that $\bttU = \bttG = \Kuchar \neq \bttC = \bttD$. 

\m 

\n See e.g.\ \cite{ATU93} for direct construction of classical Dirac observables for Minisuperspace.

\subsection{Nontrivially weak observables}

\n Reduced versus indirect makes a clear difference here, since the indirect case has a longer string of constraints in its PI. 
Only at least partly indirectly formulated case has any space for nontrivially weak such: if all constraints are reduced out, no PI is left. 

\m 

\n{\bf Example 1} Weak translational observables (a type of gauge = \K observables). 
1 particle in 1-$d$ with inhomogeneous term $W \, {\cal P} = W \, p$ supports the PI   
\be 
\sbiG^{\sw} = - W \, q \, p   \m .  
\label{PWO-11}
\ee
\n For 2 particles in 1-$d$, the inhomogeneous term $W \, {\cal P} = W (p_1 + p_2)$ supports the PI 
\be 
\sbiG^{\sw}  \es - ( a \, q_1 + b \, q_2 ) \, ( p_1 + p_2 ) 
             \es - ( a \, q_1 + b \, q_2 ) \, {\cal P}       \mma  
a + b        = W                                             \m .  
\label{Sym-PI}
\ee 
Per fixed $W$, this gives a $\mathbb{R}$ of solutions, corresponding to viewing $a$ as free.
Considering all $W$, we have a $\mathbb{R} \times \mathbb{R}_*$ of properly weak solutions. 

\m 

\n These are now however much less numerous than the strong solutions. 
This is since the strong solutions now comprise the ${\cal C}(\mathbb{R}^2)$ of suitably-smooth functions of $x_1 - x_2$ and $p_1 - p_2$. 
This is the general situation for enough degrees of freedom: that    the weak observables'   parameter space is an appendage of measure zero relative to 
                                                             that of the strong observables' function  space.
For translation-invariant mechanics, $N = 2$ particles is minimal to exhibit this effect.  
Translational mathematics being the simplest nontrivial Configurational Relationalism considered in the current article, 
this example has further senses in which it is `the simplest nontrivial example' of its kind.  

\m 

\n Looking at general particle number and dimension, we find that our method extends. 
This serves to show that \cite{DO-1} for translation-invariant models, no configuration-geometrical or pure-momentum weak observables are supported. 

\m 

\n{\bf Example 2} Weak Chronos observables.  
Now 1 particle in 1-$d$ supports the PI 
\be
\sbiC^{\sw}   \es  - W \, \frac{x}{p}  \left( \frac{p^2}{2} - k \right)
              \es  - W \, \frac{x}{p}              \, \Chronos             \m . 
\label{WCPI-11}	   
\ee
This is one function per value of $W$, or $\mathbb{R}_*$ functions in total.

\m 
	
\n For 2 particles in 1-$d$,  
\be 
\sbiC^{\sw}  \es  - \left\{ a \, \frac{q_1}{p_1}  \m + \m  b \, \frac{q_2}{p_2}  \right\}  \left( \frac{{p_1}^2  +  {p_2}^2}{2}  \m - \m  k \right) 
             \es  - \left\{ a \, \frac{q_1}{p_1}  \m + \m  b \, \frac{q_2}{p_2}  \right\} \, \Chronos                                                \mma 
a + b   =  W                                                                                                                                         \m .  
\ee  

The space of these again coincides with the corresponding translational problem. 
2 particles in 1-$d$ is again minimal for weak observables to be of zero measure relative to strong observables.
Our solution again extends to arbitrary particle number and dimension \cite{DO-1}. 

\m 

\n{\bf Example 3} Weak Dirac observables  are not just weak Chronos restricted by $\Gauge$ or vice versa. 
This is since in the $\Dirac^{\sw}$ system, firstly, the Chronos observables equation  includes a $\Gauge$    inhomogeneous term as well as a $\Chronos$ one.
Secondly,                                            the gauge observables   equations include  a  $\Chronos$ inhomogeneous term as well as   $\Gauge$   ones.  
In contrast,     the $\Chronos^{\sw}$ system has just a $\Chronos$ inhomogeneous term, 
             and the $\Gauge^{\sw}$   system has just a $\Gauge$   one.
This shows that weak observables systems do not involve a simple restriction hierarchy like strong observables ones do. 

\m 

\n{\bf Example 4} 
\be 
\w{\CanObs}^{\sw}  \m \neq \m  \CanObs^{\sw}
\ee 
is shown to be possible in \cite{DO-1}. 
A simple argument for this is that $\CanObs$ has more scope for PI terms than $\w{\CanObs}$ does, by being naturally associated with a larger constraint algebra. 
(The number of PI terms is $c^2$ for $c := \mbox{dim}(\FrC)$, since there are $c$ weak observables equations, each of which has $c$ inhomogeneous terms.)

\m 

\n{\bf Example 5}  The reduced treatment of translational RPM for 2 particles in 1-$d$ gives 
\be 
\w{\sbiD}^{\sw}   
%
%
\es  - W \, \frac{x_1 - x_2}{p_1 - p_2} \left( \frac{(p_1 - p_2)^2}{2} - k \right)       \m .
\ee
This is not however enough to have weak observables be of measure zero relative to strong observables, since we now have two constraints to two degrees of freedom. 
We do however have a general particle number and dimension solution to both the translational and chronos problems, however. 
So it is not hard to give the 3-particle, 1-$d$ minimal example of this effect. 

\m 

\n{\bf Remark 1} The usual relational numerology \cite{AMech, Minimal-N} readily lets us pick out minimal examples for strong observables dominance for further transformations 
(rotations, dilations, Euclidean, similarity, affine...) 

\m

\n{\bf Example 6}  {\it Minisuperspace} Here the sole weak observables brackets equation is 
\be
\mbox{\bf \{} \,    \scH    \mbox{\bf ,} \, \uo{\Dirac}^{\sw}     \, \mbox{\bf \}}  \es  \uo{W} \scH                                                 \m , 
\ee
giving the weak observables PDE 
\be 
\frac{\pa \,  \scH}{\pa \,  \biQ} \frac{\pa \,  \uo{\Dirac}^{\sw} }{\pa \,  \biP}  \m - \m  \frac{\pa \,  \scH}{\pa \,  \biP}  \frac{\pa \,  \uo{\Dirac}^{\sw} }{\pa \,  \biQ}  \es  \uo{W} \scH  
\ee 
\be 
\mbox{for } \m \scH  \mma  \frac{\pa \,  \scH}{\pa \,  \biQ} \m \mbox{ and } \frac{\pa \,  \scH}{\pa \,  \biP} \m \mbox{ knowns} \m .  
\ee 
\n{\bf Remark 2} Studying just minisuperspace, however, leaves one unaware of most of the diversity of types of observables, 
and of almost every effect described in the current section.  
Flat geometry and RPMs thereupon are thus rather more instructive in setting up a general theory of observables.

\section{Field Theory counterpart}\label{F-Obs} 

\subsection{Brackets algebra level} 

\n Overall, we have the following finite--field portmanteau brackets equation, 
\be
\mbox{\Large S}_{\sfA}
\left\{
\frac{\partional \, \sbcC}{\partional \, \fQ^{\sfA}}\frac{\partional \, \bea}{\partional \fP_{\sfA}} - 
\frac{\partional \, \sbcC}{\partional \, \fP_{\sfA}}\frac{\partional \, \bea}{\partional \fQ^{\sfA}}
\right\}                                                                                   \peq  0 \m .
\label{Beables-PDE}
\ee
This is to be interpreted as a $\partional$DE system, 
i.e.\ a portmanteau of  a  PDE (III.87) in the finite case 
                     and an FDE (functional differential equation)\footnote{Here $\pa \,  \bxi$ and $\pa \bxi$ are smearing functions; see Appendix A for details.}
\be
\int \d^n z \, \sumA
\left\{
\frac{\updelta(\sbcC | \pa \,  \bxi)}{\updelta \, \mQ^{\sfA}(z)}  \frac{\updelta (\uo{\bttO} | \pa \,  \uo{\bxi})}{\updelta \, \mP_{\sfA}(z)} - 
\frac{\updelta(\sbcC | \pa \,  \bxi)}{\updelta \, \mP_{\sfA}(z)}  \frac{\updelta (\uo{\bttO} | \pa \,  \uo{\bxi})}{\updelta \, \mQ^{\sfA}(z)}
\right\}  \peq   0                                                                                                                               \m .
\label{Beables-TRi-FDE}
\ee

\subsection{FDE level} 

\n{\bf Remark 1} We can moreover take 
\be 
\frac{\updelta \, \sbcC}{\updelta \, \bfQ} \m \mbox{ and } \m  \frac{\updelta \, \sbcC}{\updelta \, \bfP} \m \mbox{ to be knowns} \m , 
\ee 
leaving us with a homogeneous-linear first-order FDE system. 
We would also fix particular smearing functions or formally do not smear in locally posing and solving our FDE system.  
					 
\m 

\n{\bf Remark 2} The general form -- analogous to (III.87) for the corresponding Finite Theory PDE case -- 
\be 
\ma^{A}  \left[  \u{x}, \, y_{B}(\u{x}), \, \Phi[y_{B}(\u{x})] \right]  {\cal D}_{y_{A}} \Phi  \es  
\mb^{A}       (  \u{x}, \, y_{B}(\u{x}), \, \Phi)
\ee 
covers both our homogeneous strong observables FDE system $(\mbox{\bf b} = 0)$ and our inhomogeneous weak observables FDE system $(\mbox{\bf b} \neq 0)$.

\subsection{Flow method transcends to Banach space}

\n{\bf Remark 1} This is as far as we detail in the current series, though transfer to the tame Fr\'{e}chet space setting is also possible.  

\m 

\n{\bf Structure 8} Let ${\cal B}$ be a Banach manifold and $v$ a vector field thereupon. 
Curves, tangent vectors and tangent spaces remain defined on ${\cal B}$ \cite{AMP}.   

\m 

\n{\bf Definition 1} An {\it integral curve} of ${\cal B}$ is a curve $\gamma$ such that at each point $b$ the tangent vector is $v_b$.

\m 

\n{\bf Definition 2} The {\it differential system} on ${\cal B}$ defined by $v$ still takes its usual form,  
\be 
\dot{\phi}(\nu) = v(\phi(\nu)) \m . 
\ee
\n{\bf Remark 2} We can append a multi-index on $\phi$ and $v$ if needs be, to cover multi-component field and multi-field versions.

\m 

\n{\bf Remark 3} DE Existence and uniqueness theorems carry over \cite{AMP}.  

\m 

\n{\bf Remark 4} This guarantees a local flow (which is as much as the current Series' considerations cater for). 
As in the finite case, this provides a 1-parameter group. 

\m 

\n{\bf Structure 9} Differential forms carry over to (sufficiently smooth) ${\cal B}$.  
So do pullback, exterior differential operator and internal product \cite{AMP}. 

\m 

\n The familiar `Cartan's magic formula' for the {\bf Lie derivative} is consequently available \cite{AMP}. 

\m 

\n Thus in turn {\bf Lie dragging} remains available, as does {\bf Lie correcting} (toward implementing Configurational Relationalism) 
and the diffeomorphism interpretation of the flow.  

\m 

\n {\bf Banach Lie algebras}, and {\bf Banach Lie groups}, are well-established \cite{H72}, enabling Configurational Relationalism. 

\m 

\n Frobenius' Theorem -- as required for Lie's integral method for geometrical invariants, 
                         its canonical physics generalization and 
                         the uplifts of each of these to finding function spaces of observables thereover -- carries over as well \cite{Lang95}. 

\m 

\n{\bf End-Remark 1} All in all, local Lie Theory, 
in the somewhat broader sense required for A Local Resolution of the Problem of Time is thus established in the Banach space setting which can be taken to underlie much of Field Theory.

\subsection{Examples}

\n For conventional Gauge Theory, the observables equation imposes gauge invariance at the level of configuration space based on both space 
                                                                                                                                  and internal gauge space. 
																																  
\m 

\n In each of the first two examples below, 
\be 
\Dirac   =   \Kuchar 
         =   \gauge 
	   \neq  \unres  \m ,
\ee  
since these just have first-class linear constraints which are gauge constraints. 

\m 

\n{\bf Example 1)} Electromagnetism has the abelian algebra of constraints (VII.37). 
$\Gaugeo = \Kuchar$ for Electromagnetism solve the brackets equation 
\be
\mbox{\bf \{} \, ( \, \scG \, | \, \pa \,  \xi \, ) \mbox{\bf ,} \, ( \, \uo{\Kuchar} \, | \,  \pa \,  \uo{\bchi} \, ) \, \mbox{\bf \}}  \peq  0 
\m \m \Rightarrow \m \mbox{ the FDE }
\ee 
\beq
\underline{\pa} \cdot \frac{\updelta \Kuchar}{\updelta \underline{\mA}}  \peq  0  \m .
\eeq
This is solved by the electric and magnetic fields, 
\be 
\underline{\mE} \m \mbox{ and } \m \underline{\mB}  \es  \underline{\pa} \cr \underline{\mA}  \m , 
\ee 
and thus by a functional 
\be
{\cal F}[\underline{\mB}, \, \underline{\mE}]
\ee 
by Lemma 3.
These are not however a {\sl conjugate} pair. 
Since this looks to be a common occurrence in further examples, let us introduce the term `associated momenta' to describe it.  

\m 

\n ${\cal F}[\underline{\mB}, \, \underline{\mE}]$ can also be written in the integrated version in terms of fluxes: 
\be
{\cal F}
\left[
\iint_{\sS}\underline{\mB}\cdot \d\underline{\mS} \mma  \iint_{\sS} \underline{\mE}\cdot \d\underline{\mS}
\right] 
\es  {\cal F}[W(\upgamma), \Phi_{\sS}^{\sE}]
\ee
for electric flux $\Phi_{\sS}^{\sE}$ and loop variable
\beq
W_{\sA}(\upgamma)  \:=  \mbox{exp} \left( \, i \oint_{\upgamma} \d \u{x} \cdot \u{\mA}(x) \, \right)  \m .  
\label{Wilson-em}
\eeq
This is by use of Stokes' Theorem with $\upgamma := \pa \,  \mS$ and subsequent insertion of the exponential function subcase of Lemma 3. 
This ties the construct to the geometrical notion of holonomy. 
Moreover, these are well-known to form an {\sl over}-complete set: there are so-called {\it Mandelstam identities} between them \cite{GPBook}.

\m 
 
\n{\bf Example 2)} Its Yang--Mills generalization has the Lie algebra of constraints (VII.39)

\m 

\n $\Gaugeo = \Kuchar$ for Yang--Mills Theory solve the brackets equation 
\be 
\mbox{\bf \{} \, ( \, \scG_I \, | \, \xi^I \, ) \mbox{\bf ,} \, ( \, \uo{\Kuchar} \, | \, \uo{\bchi} \, ) \, \mbox{\bf \}} \m \approx \m  0  \m 
\Rightarrow  \m \mbox{ the FDE } 
\ee  
\beq
\uc{\u{\fD}} \cdot \frac{ \updelta \Kuchar }{ \updelta \uc{\u{\mA}} }  \m \approx \m  0  \m .   
\eeq
This is solved by Yang--Mills Theory's generalized $\uc{\u{\mE}}$ and $\uc{\u{\mB}}$, so 
\be 
{\cal F}[ \uc{\u{\mE}}, \, \uc{\u{\mB}}]
\ee 
is also a solution.  
Once again, this can be rewritten as 
\be 
{\cal F}[ W(\upgamma), \Phi_{\sS}^{\sE} ]                                                                                                         \m ,
\ee 
now for $\lFrg$-loop variable 
\beq
W_{\sA}(\upgamma)  \:=  \mbox{Tr} \left( P \, \mbox{exp} \left( i \, g \, \oint_{\upgamma} \d \u{x} \, \uc{\u{\mA}}(x) \uc{\mg}(x) \right) \right)        \m . 
\label{Wilson-YM}
\eeq 
$\mg$ are here group generators of $\lFrg_{\sY\sM}$, $g$ is the coupling constant, and $P$ is the path-ordering symbol.  

\m 

\n{\bf Example 3} The {\it cause c\'{e}l$\grave{e}$bre} of canonical treatment of observables is GR.  
In this case, one has the spatial 3-diffeomorphisms 
\be 
Diff(\bupSigma)
\ee 
momentum constraint          $\u{\scM}$ -- linear    in its momenta -- 
and a Hamiltonian constraint $\scH$   -- quadratic in its momenta -- to commute with.

\m  

\n{\bf Example 3)} $\Kuchar$ for GR as Geometrodynamics, the brackets equation   
\beq 
\mbox{\bf \{} \, ( \, \u{\scM}  \, | \, \pa \,  \u{\mL} \, ) \mbox{\bf ,} \, (  \, \uo{\Kuchar} \, | \, \pa \,  \uo{\bchi} \, )  \, \mbox{\bf \}}  \m `=' \m  0
\m \m \Rightarrow \m \m \mbox{the FDE } \m
\left(  \, \left. \left\{  \pounds_{\pa \,  \u{\sL}} \mh_{ij} \, \frac{\updelta}{\updelta \mh_{ij}}                                                      \m + \m 
                           \pounds_{\pa \,  \u{\sL}} \mp^{ij} \, \frac{\updelta}{\updelta \mp^{ij}}  \right\} 
\uo{\Kuchar} \, \right| \, \pa \,  \uo{\bchi} \, \right)                                                                                      \peq   0   \m . 
\eeq
This corresponds to the unsmeared FDE 
\beq
2 \, \mh_{jk}  {\cal D}_i  \frac{ \updelta\Kuchar }{ \updelta \mh_{ij} }                \m + \m 
\big\{  {\cal D}_i  \mp^{lj}                                                            \m - \m 
2 \, {\delta^{j}}_i  \{  {\cal D}_e \mp^{le}                                            \m + \m  
\mp^{le}{\cal D}_e  \}  \big\} \, \frac{\updelta\Kuchar}{\updelta \mp^{lj}}  \peq  0 
\m . 
\label{GR-KB}
\eeq
In the weak case, we can furthermore discard the penultimate term. 
The $\Kuchar(\bQ)$ subcase solve  
\be
2  \, \mh_{jk}  {\cal D}_i \, \frac{\updelta\Kuchar}{\updelta \mh_{ij}}  \peq   0  \m .
\label{GR-KB-Q}
\ee
These are, formally, 3-geometry quantities `${\FrG}^{(3)}$' by (\ref{GR-KB-Q}) emulating (and moreover logically preceding) 
the `momenta to the right' ordered quantum GR momentum constraint (IV.31).   
This analogy holds for the current Series' range of finite models as well (of relevance to those with nontrivial $\bFlin$).

\m 

\n{\bf Remark 1} Explicit `basis observables' are not known in this case.  

\m 

\n On the other hand, the FDE for the $\Kuchar(\bP)$ (formally `$\Pi^{\sFrG^{(3)}}$') is   
\beq
\big\{    {\cal D}_i \mp^{lj}                                                                                 \m - \m  
           2 \, {\delta^{j}}_i  \{  {\cal D}_e \mp^{le}                                                       \m + \m  
		   \mp^{le}{\cal D}_e  \}    \big\} \, \frac{ \updelta\Kuchar }{ \updelta \mp^{lj} }   \m \peq \m  0  \m . 
\label{GR-KB-P}
\eeq
\n{\bf Example 4)} D's for Geometrodynamics' $\Dirac$ require an extra FDE \cite{AObs2}  
\be 
\mbox{\bf \{} \, ( \, \scH \, | \, \pa \, \mJ \, ) \mbox{\bf ,} \, ( \, \uo{\Dirac} \, | \, \pa \,  \uo{\bzeta} \, ) \mbox{\bf  \}}  \m \peq \m  0  \m 
\Rightarrow \m \mbox{ the FDE } \m 
\ee
\beq
\left( \, 
\left.
\left\{ \, 
\frac{ \updelta \uo{\Dirac} }{ \updelta \u{\mp} }
\left\{ \, 
\u{\mG} - \u{\u{\mM}} \, \u{\cal D}^2  
\, \right\}  \m - \m 
\frac{ \updelta \uo{\Dirac} }{ \updelta \u{\mh} } \, \u{\u{\mN }} \m \u{\mp} 
\, \right\}
\pa \,  \mJ \, \right| \, \pa \,  \uo{\bzeta} 
\right)                                                                      \peq  0  \m .  
\label{42}
\eeq
This features the DeWitt vector quantities 
\beq
\underline{\mG}    \:=    \mbox{$\frac{2}{\sqrt{\sh}}$}   \big\{ \mp^{ia}  {\mp_a}^j  
                 \m - \m  \mbox{$\frac{\sp}{2}$}\mp^{ij}  \big\} 
                 \m - \m  \mbox{$\frac{1}{2\sqrt{\sh}}$}  \big\{ \mp^{ab}  \mp_{ab}   
				 \m - \m  \mbox{$\frac{\sp^2}{2}$}        \big\} \mh^{ij}
                 \m - \m  \mbox{$\frac{\sqrt{\sh}}{2}$}   \big\{ \mh^{ij}  {\cal R}        
                 \m - \m  2 \,  {\cal R}^{ij}  \big\}                                
				 \m + \m   \sqrt{\mh} \, \slLambda \, \mh^{ij}                       \m . 
\eeq
These are already familiar from the ADM equations of motion \cite{ADM}, and also $\u{\u{{\cal D}}}^2$ with components ${\cal D}^i{\cal D}^j$. 
In unsmeared form, (\ref{42}) is the FDE 
\beq
\left\{ \,  \u{\mG}  - \u{\u{\mM}} \m \u{\cal D}^2 \, \right\} \frac{ \updelta \Dirac }{ \delta \u{\mp} }  \m \peq \m   
      2 \,  \u{\mp} \m \u{\u{\mN}} \m                          \frac{ \updelta \Dirac }{ \delta \u{\mh} }  \m .
\eeq
$\biM$ and $\biN$ here are the DeWitt supermetric and its inverse respectively.

\subsection{Genericity in Field Theory}

\n In Field Theory, for instance Electromagnetism   has one observables equation, 
                         whereas Yang--Mills Theory has $g := \mbox{dim}(\lFrg_{\sY\sM})$ such. 

\m 

\n GR-as-Geometrodynamics has four observables equations: commutation with the 3 components of the momentum constraint and with the single Hamiltonian constraint.

\m 

\n So, on the one hand, for Yang--Mills Theory and for GR, PDE system genericity can be relevant. 
On the other hand, for Finite Theories, PDE system genericity is in general obscured by geometrical genericity.

\section{Conclusion}\label{Conclusion}

\subsection{Summary}

For Finite Theories, strong observables are to be found by solving brackets equations which can be recast as homogeneous linear first-order PDE systems.
Cases with a single such equation -- corresponding to a constraint (sub)algebra with a single generator -- are mathematically standard; 
the Flow Method readily applies.  
For nontrivial systems of such PDEs, there is no universal approach; the outcome depends, rather, on determinedness status.  
For observables PDE systems, moreover, over-determinedness is vanquished by integrability conditions guaranteed by Frobenius' Theorem.  
We furthermore provide firm grounding for a free alias natural \cite{CH1} characteristic problem treatment being appropriate for observables, 
indeed embodying the Taking of Function Spaces Thereover. 
Sequential use of the Flow Method then produces the requisite strong observables.  
The above workings amount to a minor extension of Lie's Integral Approach to Geometrical Invariants. 
Firstly, the observables are not just `the invariants' but rather any suitably smooth function of the invariants. 
Secondly, this extension generally plays out in phase space for us rather than in Lie's purely geometric setting.  

\m 

\n Properly weak observables obey the inhomogeneous-linear counterpart of the above. 
Their general solution thus splits into complementary function plus particular integral, 
which roles are played by strong and nontrivially-weak observables respectively. 
For examples with enough degrees of freedom, moreover, properly weak observables give but a measure-zero extension to the space of strong observables. 

\m 

\n We illustrate all of strong versus properly weak observables, 
of unrestricted, Dirac and middlingly restricted observables, 
as well as of properly weak observables' dependence on extent of reduction.  
Also fully reduced treatments have no room for properly weak observables. 
This and the above zero-measure comment are two reasons to not place too much stock in further developing the theory of properly weak observables.  

\m 

\n The above summary corresponds to the first major extension of \cite{ABook}'s treatment of Background Independence and the Problem of Time. 
This is by providing a concrete theory of solving for observables on the endpoint of the first branch of Fig 1.b), 
for now in the local, classical finite-theory setting.  

\m 

\n For Field Theories, observables equations give instead linear first-order functional differential equation systems.  
Banach (or tame Fr\'{e}chet) Calculus is however sufficiently standard that the Flow Method and free characteristic problems still apply. 
These calculi support the Lie-theoretic combination of machinery that our Local Resolution of the Problem of Time requires.
In this way, Field Theory -- exemplified by Electromagnetism and Yang--Mills Theory in the current Article -- and GR are included in our 
Local Resolution of the Problem of Time's formulation.

\subsection{Present and future of the pedagogy of observables}

Pedagogical difficulties with presenting observables moreover abound. 
Many quantum treatises, as well as popular accounts \cite{W-Obs}, immediately launch into the quantum version.
They do not mention the corresponding classical counterpart because of its relative simplicity.   
In the process, they miss the point that the classical version becomes nontrivial in the presence of constraints. 
This difficulty moreover recurs at the quantum level, where it is missed out again. 
In this way, the most advanced theory of observables -- of quantum constrained observables -- is not mentioned on Wikipedia \cite{W-Obs}. 

\m 

\n The suggested restructuring is to start with classical unconstrained functions over phase space (and configuration space, with connections to pure geometry). 
Then, on the one hand, one is to consider the nontrivialities of classical constrained systems. 
On the other hand, one is to keep the existing quantum account as the other main complicating factor. 
This is now to emphasize that only small subalgebras of the classical observables algebra can be consistently quantized. 
Finally, these two sources of complexity are to be combined. 
Article X spells out further sources of improvement, 
along the lines of the current paragraph's canonical notions of observables having spacetime, path or histories observables analogues. 
The common theme is that of Function Spaces Thereover, meaning over state spaces. 
Be that (un)constrained classical phase space, the space of spacetimes or some quantum state space.

\subsection{The theory of strong observables has at least a presheaf flavour}

Taking Function Spaces Thereover models moreover multiplicity of function spaces over state spaces. 
This is in the sense of applying various different consistent constraint algebraic substructures.  
For strong observables -- the generic part of observables theory for sufficient degrees of freedom -- 
these combine in the manner of applying successive restrictions to the observables DE's characteristic surfaces as per Theorem 3.  
Such inter-relation by restriction maps amounts to Taking Presheaves Thereover.  
How widely this can be extended to Taking Sheaves Thereover remains to be determined. 
This is an interesting question to investigate, due to sheaves possessing further localization and globalization properties;\footnote{See in particular 
\cite{Ghrist, Wells, Wedhorn}         for introductory accounts of sheaves, 
\cite{Sheaves1, Sheaves2, Hartshorne} for more advanced accounts, and 
\cite{FH87}                           for a pioneering paper on the use of presheaves to model quantum observables.}
it is however beyond the reach of the current Series' {\sl local} treatise.

\begin{appendices}

\section{TRi observables}\label{TRi-Obs}
%
{            \begin{figure}[!ht]
\centering
\includegraphics[width=1.0\textwidth]{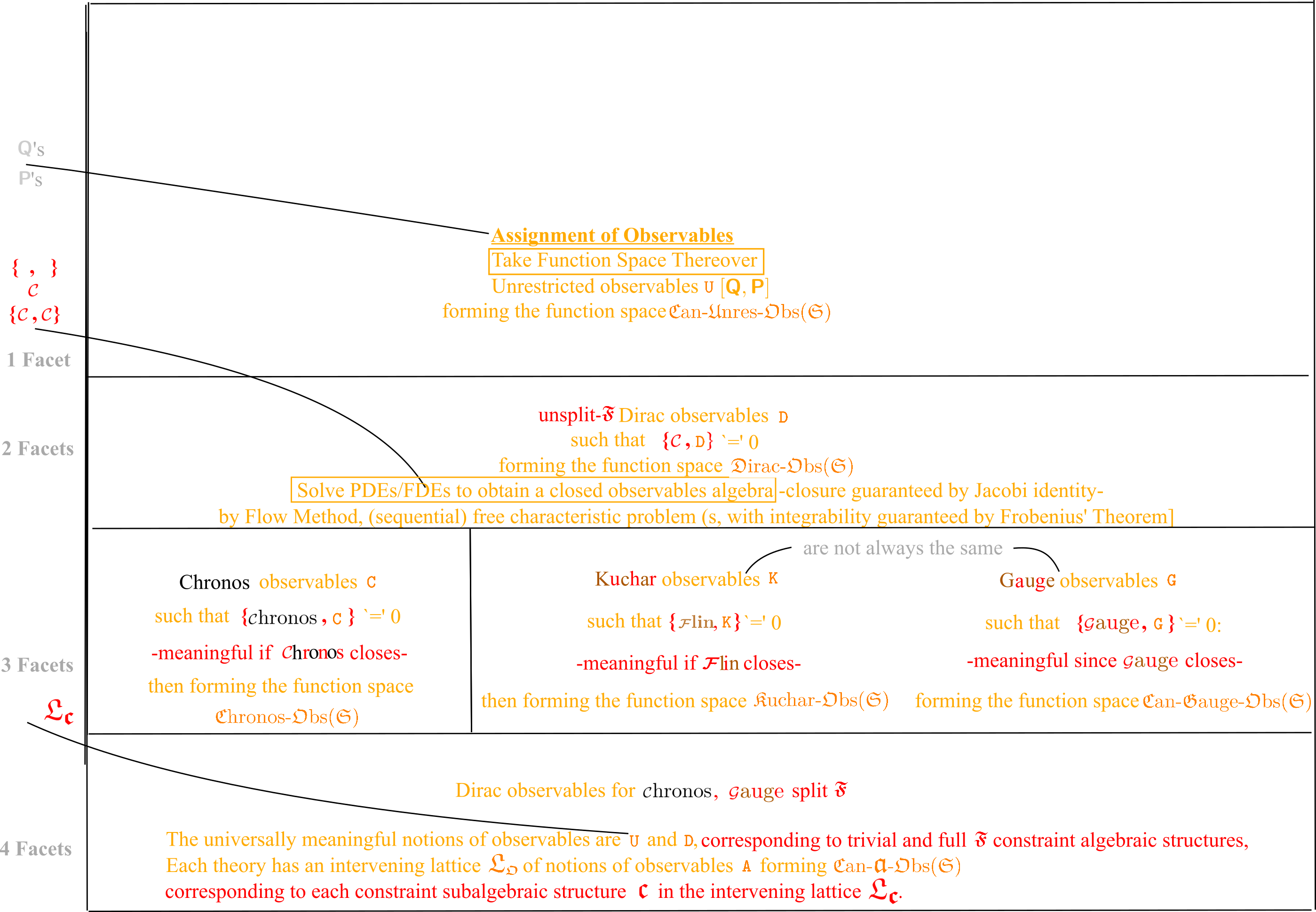} 
\caption[Text der im Bilderverzeichnis auftaucht]{    \footnotesize{This continuation of the `technicolour guide' to the Problem of Time 
sits to the left of Fig VII.5.  
The new Assignment of Observables content is highlighted in orange.} }
\label{FST-EitoB-Interference} \end{figure}      }

\n 1) $\biQ$, $\biP$ and $\Phase$ are already-TRI.  
So are $\sbcC$ and $\mbox{\bf \{} \m \mbox{\bf ,} \, \m \mbox{\bf \}}$, and thus the definition of constrained observables as well.  

\m 

\n 2) The current article's smearing functions are given in TRi form since first-class constraints are both trivially weak observables and TRi-smeared, 
pointing to all observables requiring TRi-smearing as well.  

\m 

\n 3) Observables algebraic structures are already-TRi in the Finite Theory case, or readily rendered TRi by adopting TRi-smearing in the Field Theory case.  

\m 

\n 4) Split $\sbcC$ into Temporal and Configurational Relationalism parts has the knock-on effect (already in III) 
that some notions of observables are just Configurationally Relational -- \K observables -- 
                                  or just Temporally        Relational -- Chronos observables -- in theories admitting such a split. 
GR permits \K observables but not Chronos observables, whereas RPM supports both.

\m 

\n 5) See Fig \ref{FST-EitoB-Interference} for Expression in Terms of Observables' further Problem of Time facet interferences.  
This scantness of interaction is testimony to the great decoupling of facets alluded to in the Introduction.  
Further historical problems with the Problem of Observables stem from failing to distinguish between canonical and spacetime notions of observables.   
Discussing these must however await detailed consideration of the spacetime observables version in Article X.

\end{appendices}
  

\end{document}